\documentclass[%
reprint,
nofootinbib,
 amsmath,amssymb,
 aps,
]{revtex4-2}
\usepackage{stmaryrd}

\usepackage{graphicx}% Include figure files
\usepackage{dcolumn}% Align table columns on decimal point
\usepackage{bm}% bold math
\usepackage{hyperref}% add hypertext capabilities
\usepackage[caption=false]{subfig}
\usepackage{graphicx}
\begin{document}

\preprint{APS/123-QED}

\title{Searching for gravitational wave burst in PTA data with piecewise linear functions}
\author{Heling Deng}
\author{Bence Bécsy}
\author{Xavier Siemens}
\affiliation{Department of Physics, Oregon State University, Corvallis, OR 97331, USA}
\author{Neil J. Cornish}
\affiliation{eXtreme Gravity Institute, Department of Physics, Montana State University, Bozeman, MT 59717, USA}
\author{Dustin R. Madison}
\affiliation{Department of Physics, University of the Pacific, 3601 Pacific Avenue, Stockton, CA 95211, USA}
\begin{abstract}

{Transient gravitational waves (aka gravitational wave bursts)} within the nanohertz frequency band could be generated by a variety of astrophysical phenomena such as the encounter of supermassive black holes, the kinks or cusps in cosmic strings, or other as-yet-unknown physical processes. Radio-pulses emitted from millisecond pulsars could be perturbed by passing gravitational waves, hence the correlation of the perturbations in a pulsar timing array can be used to detect and characterize burst signals with a duration of $\mathcal{O}(1\text{-}10)$ years. We propose a fully Bayesian framework for the analysis of the pulsar timing array data, where the burst waveform is generically modeled by piecewise straight lines, and the waveform parameters in the likelihood can be integrated out analytically. As a result, with merely three parameters (in addition to those describing the pulsars' intrinsic and background noise), one is able to efficiently search for the existence and the sky location of {a burst signal}. If a signal is present, the posterior of the waveform can be found without further Bayesian inference. We demonstrate this model by analyzing simulated data sets containing a stochastic gravitational wave background {and a burst signal generated by the parabolic encounter of two supermassive black holes.}

\end{abstract}

\maketitle

\section{Introduction}

Millisecond pulsars are highly magnetized rotating neutron stars with periods $\mathcal{O}(1\text{-}10)$ milliseconds. Beams of electromagnetic radiation emitting from the magnetic poles rotate about the star's spinning axis and may hit us as radio-pulses once per period. Known to have very stable rotations, millisecond pulsars are highly sensitive probes of their environments, including gravitational waves (GWs). These waves cannot be inferred by observing a single pulsar, but correlations among an array of pulsars can in principle be hunted down. In particular, the detection of Hellings \& Downs (HD) correlations would be an unambiguous signature of a stochastic gravitational wave background (SGWB) \cite{Hellings:1983fr}. Pulsar timing array (PTA) observations so far typically have a sampling interval of weeks and span over $\mathcal{O}(10)$ years, implying a sensitive GW frequency range of around $1\text{-}100$ nHz.

{Recently, the North American Nanohertz Observatory for Gravitational Waves (NANOGrav) \cite{McLaughlin:2013ira} reported the first convincing evidence for a nHz SGWB in our universe \cite{NANOGrav:2023gor}. The analysis of the NANOGrav 15-yr data set shows a red noise process that has a spectrum common among all pulsars and that is spatially correlated among pulsar pairs in a manner consistent with HD correlations.}

A promising source of SGWB in PTA is the combined emission from an ensemble of inspiraling supermassive black hole (SMBH) binaries. Most galaxies have an SMBH with mass $10^{6}\text{-}10^{10}M_{\odot}$ sitting at the center \cite{Kormendy:2013dxa}. When two galaxies merge, two SMBHs may find each other and form a binary, emitting gravitational radiation for a time span much larger than the PTA observation period. {The investigation of constraints on SMBH binaries from the NANOGrav 15-yr data set can be found in Ref. \cite{NANOGrav:2023hfp}.} An SGWB in the nHz band could also be generated by physics in the early universe, such as inflation \cite{Grishchuk:1974ny,Starobinsky:1979ty,Rubakov:1982df,Fabbri:1983us}, phase transitions \cite{Kosowsky:1991ua,Kosowsky:1992rz,Kosowsky:1992vn,Kamionkowski:1993fg,Caprini:2007xq,Huber:2008hg,Hindmarsh:2013xza,Giblin:2013kea,Giblin:2014qia,Hindmarsh:2015qta,NANOGrav:2021flc} and cosmic strings \cite{Vilenkin:1981bx,Damour:2004kw,Buchmuller:2020lbh,Ellis:2020ena,Blanco-Pillado:2021ygr,Hindmarsh:2022awe}. {Up-to-date PTA constraints on new physics can be found in Refs. \cite{NANOGrav:2023hvm,Antoniadis:2023xlr,Smarra:2023ljf}.}

Besides the stochastic background, events with strong GW emissions from certain sky locations could be detected individually. An example under active search is continuous waves emitted from the brightest SMBH binary, the detection of which would provide direct evidence of the existence of SMBH binaries \cite{Sesana:2008xk,Rosado:2015epa,Kelley:2017vox,Becsy:2022pnr}. {Several  searches have been carried out over the years, setting increasingly stringent upper bounds on these sources \cite{Yardley:2010kv,NANOGrav:2014zwv,Zhu:2014rta,Babak:2015lua,Aggarwal:2018mgp,NANOGrav:2023bts,IPTA:2023ero,NANOGrav:2023pdq,Antoniadis:2023aac}. }

In this paper, we shall focus on searching for the strongest GW burst {(or GW transient)} with a duration comparable to the PTA observation period. Bursts of this kind could come from a variety of physical processes: the encounter of two SMBHs; cosmic string cusps or kinks \cite{Damour:2000wa,Damour:2001bk,Yonemaru:2020bmr}; the GW memory effect, i.e., a permanent deformation of spacetime after a violent event, such as the merger of two SMBHs (the merger itself also emits GWs, but they are not in the nHz band) \cite{Cordes:2012zz,Madison:2014vca,vanHaasteren:2009fy,NANOGrav:2015xuc,Wang:2014zls,NANOGrav:2019vto,Sun:2022hvp}. Furthermore, there may exist other phenomena generating bursts with unknown waveforms. 

A number of methods of searching for and reconstructing a generic burst signal in the PTA data have been proposed and developed over the years.\footnote{Burst-search around the kilohertz frequency band has been studied extensively in the context of ground-based interferometric GW detectors such as LIGO and Virgo \cite{KAGRA:2021tnv,KAGRA:2021bhs,Szczepanczyk:2022urr}.} In Ref. \cite{Finn:2010ph}, an analytical hybrid (frequentist-Bayesian) approach was adopted, where each data point was assigned a parameter characterizing the burst waveform, and a maximum a posteriori estimate was performed to fix the hyperparameters. This method was improved in Ref. \cite{Deng:2014kua} by a Bayesian nonparametric analysis. In Refs. \cite{Zhu:2014rta,Zhu:2015tua,Madison:2015txa}, frequentist frameworks were developed (in the time domain or the frequency domain), where piecewise linear functions were introduced to describe the burst waveform, and a least-squares fitting process was performed to obtain the estimates of the waveform parameters. Recently, a Bayesian algorithm was implemented in Refs. \cite{Becsy:2020utk,Becsy:2023}, where a generic burst is modeled by the superposition of Morlet-Gabor wavelets.

In the present work, we present an efficient Bayesian method to search for the strongest signal from a burst with an unknown waveform in PTA data. Motivated by Ref. \cite{Madison:2015txa}, we model the burst waveform (with two polarizations) with two piecewise linear functions in the time domain. In our model, the existence and the sky location of the burst can be determined with only three parameters (in addition to those describing the pulsars' intrinsic and common noise), \emph{with the parameters characterizing the burst waveform analytically integrated out.} If a signal is indeed present in the data, its waveform can then be straightforwardly extracted without performing further Bayesian inference. We shall test this method by analyzing simulated PTA data sets that contain an SGWB.

The rest of the paper is organized as follows. In Sec. \ref{waveform approximation} we describe how the burst waveform can be modeled by piecewise linear functions, which leads to a simple expression of the marginalized likelihood discussed in detail in Sec. \ref{likelihood}. In Sec. \ref{likelihood} we also show how to detect the burst's sky location and reconstruct the waveform. The model will be demonstrated by the analyses of three simulated data sets in Sec. \ref{simulated data}. Conclusions are summarized and discussed in Sec. \ref{discussion}

\section{Waveform modeling\label{waveform approximation}}

In pulsar timing, the times of arrival (TOAs) of radio-pulses from a millisecond pulsar are measured and compared with predictions based on a timing model that describes the pulsar physics (e.g., the spin period, the spin-down rate, {etc.}). The differences are called timing residuals. While they mainly come from deviations in the timing model, white noise from measurement uncertainties and red noise from the pulsar's intrinsic instabilities, the residuals may also be disturbed by passing GWs, such as a possible stochastic background or deterministic signals from certain physical phenomena. In this section, we describe the residuals of a pulsar induced by a generic GW burst and introduce our model, where the waveform of the burst is approximated by piecewise straight lines. The residuals can then be expressed in a simple form.

\subsection{Timing residuals induced by burst}

The location of the observer is set as the Solar System barycenter (SSB), which sits at the origin of Cartesian coordinates defined by orthonormal vectors $(\hat{x},\hat{y},\hat{z})$. {The north celestial pole is in the $\hat{z}$-direction and the vernal equinox is in the $\hat{x}$-direction.} The sky location of the GW burst can then be determined by the polar and azimuthal angles $(\theta,\phi)$. Three useful orthonormal vectors are
\begin{align}
\hat{\Omega}&=-\sin\theta\cos\phi \hat{x}-\sin\theta\sin\phi\hat{y}-\cos\theta\hat{z},\\
\hat{m}&=-\sin\phi\hat{x}+\cos\phi\hat{y},\\
\hat{n}&=-\cos\theta\cos\phi\hat{x}-\cos\theta\sin\phi\hat{y}+\sin\theta\hat{z},
\end{align}
where $\hat{\Omega}$ points from the GW source to SSB, and $\hat{m}$ and $\hat{n}$ are vectors that are useful for describing the two polarization tensors of the source. 

{It can be shown that the GW brings two redshifting signatures to a pulsar's TOAs: perturbations to the timing residuals when the wave reaches the pulsar (leading to the ``pulsar term'') and when it reaches the Earth (the ``Earth term'') \cite{estabrook1975response}.} Since the duration of a GW burst of interest ($\sim1\text{-}10$ years) is much smaller than the time it takes for a radio-pulse to travel from the pulsar to the Earth (thousands of years), it is unlikely that a pulsar's Earth term and pulsar term are both present in a PTA data set; it is also unlikely that pulsar terms from different pulsars are correlated. On the other hand, all Earth terms show up within the same period. It is thus safe to neglect the pulsar terms. 

The residuals of a single pulsar induced by a burst can then be written in the following form
\begin{equation}
h(t)=F^{+}(\hat{\Omega})H^{+}(t)+F^{\times}(\hat{\Omega})H^{\times}(t).
\end{equation}
Here $H^{+,\times}(t)$ represent the perturbations to the residuals when the burst reaches the Earth at time $t$, and $F^{+,\times}$ are so-called antenna pattern functions that describe the response of an Earth--pulsar system to the GW signal, given by
\begin{align}
F^{+}(\hat{\Omega})&=\frac{1}{2}\frac{(\hat{m}\cdot\hat{p})^{2}-(\hat{n}\cdot\hat{p})^{2}}{1+\hat{\Omega}\cdot\hat{p}},\\
F^{\times}(\hat{\Omega})&=\frac{(\hat{m}\cdot\hat{p})(\hat{n}\cdot\hat{p})}{1+\hat{\Omega}\cdot\hat{p}},
\end{align}
where $\hat{p}$ is a unit vector pointing from SSB to the pulsar.

\subsection{Waveform described by piecewise straight lines}

By the previous subsection, the timing residuals of the $I$th pulsar caused by a GW burst reaching the Earth at time $t$ can be written as
\begin{equation}
h_{\tiny{I}}(t)=F_{\tiny{I}}^{+}H^{+}(t)+F_{\tiny{I}}^{\times}H^{\times}(t).
\end{equation}
If the sky location of the pulsar is known, $F_{\tiny{I}}^{+,\times}$ are functions of the sky location of the burst $(\theta,\phi)$. For certain physical processes, such as GW memory effects or cosmic string cusps, the waveform $H^{+}(t)$ and $H^{\times}(t)$ can be determined by theories. For a generic burst, however, one needs a signal model that can describe a wide variety of waveforms.

\begin{figure}
\centering
\includegraphics[scale=0.4]{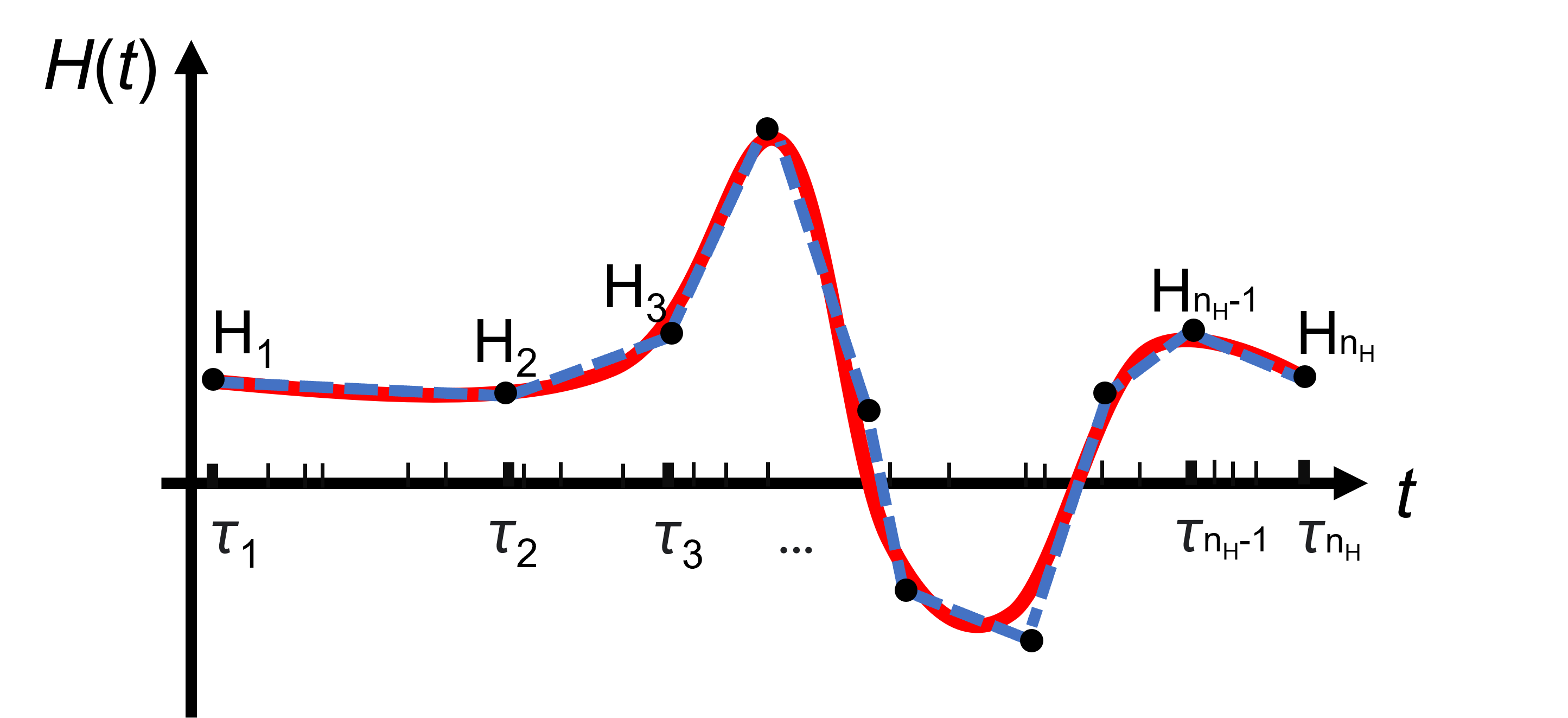}

\caption{\label{fig:PWL}Illustration of waveform $H(t)$ (red, solid) being approximated by piecewise straight lines (blue, dashed). The observation period is divided into $n_{\tiny{H}}-1$ parts, each grid point being denoted by $\tau_{\mu}$, to which $n_{\tiny{H}}$ quantities $\mathsf{H}_{\mu}$ are assigned to represent $H(\tau_{\mu})$.}
\end{figure}

We will {model} $H^{+}(t)$ and $H^{\times}(t)$ with two independent piecewise linear functions. An illustration is shown in Fig. \ref{fig:PWL}. To this end, we divide the $n_{\text{toas}}$ recorded TOAs of the pulsar into $n_{\tiny{H}}-1$ (not necessarily even) parts, where $n_{\tiny{H}}\ll n_{\text{toas}}$.  Let $\tau_{\mu}$ be the time at the $n_{\tiny{H}}$ grid points, with $\mu=1,2,...,n_{\tiny{H}}$. We assign to each $\tau_{\mu}$ two quantities: $\mathsf{H}_{\mu}^{+}$ and $\mathsf{H}_{\mu}^{\times}$. The $2n_{\tiny{H}}$ quantities $\begin{pmatrix}\mathsf{H}_{1}^{+} & \mathsf{H}_{2}^{+} & \cdots & \mathsf{H}_{n_{\tiny{H}}}^{+} & \mathsf{H}_{1}^{\times} & \mathsf{H}_{2}^{\times} & \cdots & \mathsf{H}_{n_{\tiny{H}}}^{\times}\end{pmatrix}\equiv\boldsymbol{\mathsf{H}}^{\top}$ will be the parameters characterizing the waveform: $H^{+}$ ($H^{\times}$) at any time $t$ can be estimated from the linear combination of the two neighboring quantities in $\mathsf{H}_{\mu}^{+}$ ($\mathsf{H}_{\mu}^{\times}$). For example, for $\tau_{1}<t<\tau_{2}$, $H^{+}(t)$ can be estimated as
\begin{align}
H^{+}(t)&\approx\mathsf{H}_{1}^{+}+\frac{\mathsf{H}_{2}^{+}-\mathsf{H}_{1}^{+}}{\tau_{2}-\tau_{1}}\left(t-\tau_{1}\right)\nonumber\\
&=\begin{pmatrix}\frac{\tau_{2}-t}{\tau_{2}-\tau_{1}} & \frac{t-\tau_{1}}{\tau_{2}-\tau_{1}}\end{pmatrix}\begin{pmatrix}\mathsf{H}_{1}^{+}\\
\mathsf{H}_{2}^{+}
\end{pmatrix}.
\end{align}
Let $\mathsf{t}_{k}$ be the list of TOAs, where $k=1,2,...,n_{\text{toas}}$. We have
\begin{widetext}
\begin{equation}
\begin{pmatrix}H^{+}(\mathsf{t}_{1})\\
H^{+}(\mathsf{t}_{2})\\
\vdots\\
H^{+}(\mathsf{t}_{n_{\text{toas}}-1})\\
H^{+}(\mathsf{t}_{n_{\text{toas}}})
\end{pmatrix} \approx\begin{pmatrix}\frac{\tau_{2}-\mathsf{t}_{1}}{\tau_{2}-\tau_{1}} & \frac{\mathsf{t}_{1}-\tau_{1}}{\tau_{2}-\tau_{1}} & 0 & \cdots & 0 & 0\\
\frac{\tau_{2}-\mathsf{t}_{2}}{\tau_{2}-\tau_{1}} & \frac{\mathsf{t}_{2}-\tau_{1}}{\tau_{2}-\tau_{1}} & 0 & \cdots & 0 & 0\\
\vdots & \vdots & \vdots & \ddots & \vdots & \vdots\\
0 & 0 & 0 & \cdots & \frac{\tau_{n_{\tiny{H}}}-\mathsf{t}_{n_{\text{toas}}-1}}{\tau_{n_{\tiny{H}}}-\tau_{n_{\tiny{H}}-1}} & \frac{\mathsf{t}_{n_{\text{toas}}-1}-\tau_{n_{\tiny{H}}-1}}{\tau_{n_{\tiny{H}}}-\tau_{n_{\tiny{H}}-1}}\\
0 & 0 & 0 & \cdots & \frac{\tau_{n_{\tiny{H}}}-\mathsf{t}_{n_{\text{toas}}}}{\tau_{n_{\tiny{H}}}-\tau_{n_{\tiny{H}}-1}} & \frac{\mathsf{t}_{n_{\text{toas}}}-\tau_{n_{\tiny{H}}-1}}{\tau_{n_{\tiny{H}}}-\tau_{n_{\tiny{H}}-1}}
\end{pmatrix}\begin{pmatrix}\mathsf{H}_{1}^{+}\\
\mathsf{H}_{2}^{+}\\
\vdots\\
\mathsf{H}_{n_{\tiny{H}}-1}^{+}\\
\mathsf{H}_{n_{\tiny{H}}}^{+}
\end{pmatrix} \equiv\mathsf{P}_{\tiny{I}}\boldsymbol{\mathsf{H}}^{+},
\end{equation}
\end{widetext}
where $\mathsf{P}_{\tiny{I}}$ is an $n_{\text{toas}}\times n_{\tiny{H}}$ matrix specific to the $I$th pulsar (since it depends on that pulsar's recorded
TOAs). The burst signal can then be estimated as
\begin{align}
h_{\tiny{I}}&\approx F_{\tiny{I}}^{+}\mathsf{P}_{\tiny{I}}\boldsymbol{\mathsf{H}}^{+}+F_{\tiny{I}}^{\times}\mathsf{P}_{\tiny{I}}\boldsymbol{\mathsf{H}}^{\times}\nonumber\\
&=\begin{pmatrix}F_{\tiny{I}}^{+}\mathsf{P}_{\tiny{I}} & F_{\tiny{I}}^{\times}\mathsf{P}_{\tiny{I}}\end{pmatrix}\begin{pmatrix}\boldsymbol{\mathsf{H}}^{+}\\
\boldsymbol{\mathsf{H}}^{\times}
\end{pmatrix}\equiv\mathsf{S}_{\tiny{I}}\boldsymbol{\mathsf{H}}.
\end{align}
where $\mathsf{S}_{\tiny{I}}\equiv\begin{pmatrix}F_{\tiny{I}}^{+}\mathsf{P}_{\tiny{I}} & F_{\tiny{I}}^{\times}\mathsf{P}_{\tiny{I}}\end{pmatrix}$ is an $n_{\text{toas}}\times2n_{\tiny{H}}$ matrix. Let $\boldsymbol{h}$ be a concatenated vector composed of all pulsar's residuals generated by the burst, we have
\begin{equation}
\boldsymbol{h}\approx\boldsymbol{\mathsf{S}}\boldsymbol{\mathsf{H}}.\label{eq:SA}
\end{equation}
Here we have defined $\boldsymbol{\mathsf{S}}=\begin{pmatrix}\mathsf{S}_{1}^{\top} & \mathsf{S}_{2}^{\top} & \cdots & \mathsf{S}_{n_{\text{psr}}}^{\top}\end{pmatrix}^{\top},$ where $n_{\text{psr}}$ is the number of pulsars under consideration. An advantage of using  $\boldsymbol{\mathsf{H}}$ to describe the waveform is that, compared with higher-order polynomials or Fourier series, locally bad TOAs do not contaminate the estimates of $H^{+,\times}(t)$ over large spans of data \cite{Madison:2015txa}.

\section{Likelihood \label{likelihood}}

The timing residuals induced by a burst modeled in the previous section will be used to construct the likelihood of the PTA data. It will be shown in this section that it is possible to {analytically integrate out the waveform parameters $\boldsymbol{\mathsf{H}}$.} By so doing, one is able to determine the existence and the sky location of the burst without reconstructing the waveform. Such a marginalized likelihood allows an efficient Bayesian search.

\subsection{Likelihood without deterministic signals}

PTA residuals are generated by various processes. In the absence of deterministic signals, the residuals are often modeled as the sum of contributions from {timing model deviations}, white noise and red noise (such as an SGWB) \cite{vanHaasteren:2008yh,Lentati:2012xb,vanHaasteren:2014qva,taylor2021nanohertz}:
\begin{equation}
\boldsymbol{r}=\boldsymbol{n}+\boldsymbol{M}\boldsymbol{\epsilon}+\boldsymbol{F}\boldsymbol{a}.
\end{equation}
Here $\boldsymbol{r}$ is a concatenated vector composed of all pulsar's residuals; $\boldsymbol{n}$ contains white noise from the radiometer, instrumental effects, etc.; $\boldsymbol{M}$ is the timing model's design matrix basis, and $\boldsymbol{\epsilon}$ is a vector of the corresponding coefficients; $\boldsymbol{F}$ represents the Fourier basis of the red noise, and $\boldsymbol{a}$ is a vector of the corresponding amplitudes. Since $\boldsymbol{n}$ is expected to behave as white noise, vector $\boldsymbol{r}-\boldsymbol{M}\boldsymbol{\epsilon}-\boldsymbol{F}\boldsymbol{a}$ obeys the Gaussian distribution. To simplify the notation, we shall write a zero-mean Gaussian (normal) distribution with covariance matrix $\boldsymbol{D}$ as
\begin{equation}
\mathcal{N}(\boldsymbol{x}|\boldsymbol{D})\equiv\frac{\exp\left(-\frac{1}{2}\boldsymbol{x}^{\top}\boldsymbol{D}^{-1}\boldsymbol{x}\right)}{\sqrt{\det(2\pi\boldsymbol{D})}}.
\end{equation}
The PTA likelihood is then given by
\begin{equation}
\mathcal{L}(\boldsymbol{r}|\boldsymbol{b})=\mathcal{N}(\boldsymbol{r-Tb}|\boldsymbol{N}),
\end{equation}
where $\boldsymbol{N}$ is the white noise covariance matrix, $\boldsymbol{T}=\begin{pmatrix}\boldsymbol{M} & \boldsymbol{F}\end{pmatrix}$ and $\boldsymbol{b}=\begin{pmatrix}\boldsymbol{\epsilon} & \boldsymbol{a}\end{pmatrix}^{\top}$. The prior on $\boldsymbol{b}$ can also be set as Gaussian:
\begin{equation}
\pi(\boldsymbol{b}|\boldsymbol{\eta})=\mathcal{N}(\boldsymbol{b}|\boldsymbol{B}).\label{eq:bB}
\end{equation}
Here the covariance matrix is given by $\boldsymbol{B}=\text{diag}(\infty,\boldsymbol{\phi}(\boldsymbol{\eta}))$, where $\boldsymbol{\eta}$ contains hyperparameters that control $\boldsymbol{B}$. The timing model coefficients $\boldsymbol{\epsilon}$ are well constrained by observations; their inference is likelihood-dominated. Hence we can impose on them a Gaussian prior of infinite variance. The covariance matrix of the Fourier coefficients $\boldsymbol{a}$ includes all possible intrinsic or common red noise processes:
\begin{equation}
\phi_{({\tiny{I}}i)({\tiny{J}}j)}=\left\langle a_{{\tiny{I}} {i}}a_{{\tiny{J}} j}\right\rangle =\delta_{ij}\left(\delta_{\tiny{IJ}}\varphi_{{\tiny{I}}i}+\Gamma_{\tiny{IJ}}\Phi_{i}\right),
\end{equation}
where $I,J$ range over pulsars and $i,j$ over Fourier components; $\delta_{ij}$ is the Kronecker delta; $\varphi_{{\tiny I}i}$ describes the spectrum of intrinsic red noise in pulsar $I$; and  $\Gamma_{\tiny{IJ}}\Phi_{i}$ describes processes with a common spectrum across all pulsars and inter-pulsar correlations. For an isotropic SGWB with HD correlations,  $\Gamma_{\tiny IJ}$ is the HD function of pulsar angular separations, and $\Phi_{i}$ is usually assumed to obey a power law characterized by amplitude $A$ and spectral index $\gamma$:
\begin{equation}
\Phi_{i}=\frac{A^{2}}{12\pi^{2}}\frac{1}{T}\left(\frac{f_{i}}{1\ \text{yr}^{-1}}\right)^{-\gamma}1\ \text{yr}^{-3}.\label{eq:power}
\end{equation}
Here $f_{i}$ is the frequency of the $i$th Fourier component and $T$ is the maximum TOAs extent. For an SWGB generated by inspiral SMBH binaries, $\gamma$ is expected to be $13/3$ \cite{phinney2001practical}. 

The full hierarchical PTA posterior can then be written as
\begin{equation}
p(\boldsymbol{b},\boldsymbol{\eta}|\boldsymbol{r})=\mathcal{L}(\boldsymbol{r}|\boldsymbol{b})\pi(\boldsymbol{b}|\boldsymbol{\eta})\pi(\boldsymbol{\eta}),
\end{equation}
where $\pi(\boldsymbol{\eta})$ is the hyperprior on $\boldsymbol{\eta}$. Compared with the hyperparameters in $\boldsymbol{\eta}$, e.g., the amplitude ($A$) and spectral index ($\gamma$) of the SGWB power spectrum, parameters in  $\boldsymbol{b}$, i.e.,  the design matrix coefficients $\boldsymbol{\epsilon}$ and the Fourier coefficients $\boldsymbol{a}$, are usually not of particular interest. Note that the hierarchical likelihood $\mathcal{L}(\boldsymbol{r}|\boldsymbol{b})\pi(\boldsymbol{b}|\boldsymbol{\eta})$ is a Gaussian function for $\boldsymbol{b}$. One can then integrate out these parameters analytically and obtain the marginalized likelihood that only depends on $\boldsymbol{\eta}$ \cite{Lentati:2012xb,vanHaasteren:2012hj}:
\begin{equation}
\mathcal{L}(\boldsymbol{r}|\boldsymbol{\eta})=\int\mathcal{L}(\boldsymbol{r}|\boldsymbol{b})\pi(\boldsymbol{b}|\boldsymbol{\eta})\text{d}\boldsymbol{\epsilon}\text{d}\boldsymbol{a}=\mathcal{N}(\boldsymbol{r}|\boldsymbol{C}),\label{eq:marginalized likelihood}
\end{equation}
where $\boldsymbol{C}=\boldsymbol{N}+\boldsymbol{T}\boldsymbol{B}\boldsymbol{T}^{\top}$, and we have used the Woodbury identity that gives $\boldsymbol{C}^{-1}=\boldsymbol{N}^{-1}+\boldsymbol{N}^{-1}\boldsymbol{T}\left(\boldsymbol{B}^{-1}+\boldsymbol{T}^{\top}\boldsymbol{N}^{-1}\boldsymbol{T}\right)^{-1}\boldsymbol{T}^{\top}\boldsymbol{N}^{-1}$. This is practically the likelihood implemented within production-level GW search pipelines, such as ENTERPRISE\footnote{https://github.com/nanograv/enterprise, https://github.com/nanograv/enterprise\_extensions} \cite{enterprise}.

\subsection{Marginalized likelihood including the burst}

In the previous subsection, we consider the PTA likelihood when only stochastic processes are present. When the burst is taken into account, residuals $\boldsymbol{r}$ in Eq. (\ref{eq:marginalized likelihood}) should be replaced by $\boldsymbol{r}-\boldsymbol{h}\approx\boldsymbol{r}-\boldsymbol{\mathsf{S}}\boldsymbol{\mathsf{H}}$. To simplify the notation, we define inner product $\left\llbracket \boldsymbol{x}|\boldsymbol{y}\right\rrbracket =\boldsymbol{x}^{\top}\boldsymbol{C}^{-1}\boldsymbol{y}$. The likelihood in our ``$\boldsymbol{h}=\boldsymbol{\mathsf{S}}\boldsymbol{\mathsf{H}}$'' model is then given by
\begin{align}
\mathcal{L}(\boldsymbol{r}|\boldsymbol{\eta},\theta,\phi,\boldsymbol{\mathsf{H}}) & =\mathcal{N}(\boldsymbol{r}-\boldsymbol{\mathsf{S}}\boldsymbol{\mathsf{H}}|\boldsymbol{C})\nonumber\\
 & =\mathcal{N}(\boldsymbol{r}|\boldsymbol{C})\exp\left(\left\llbracket \boldsymbol{r}|\boldsymbol{\mathsf{S}}\right\rrbracket \boldsymbol{\mathsf{H}}-\frac{1}{2}\boldsymbol{\mathsf{H}}^{\top}\left\llbracket \boldsymbol{\mathsf{S}}|\boldsymbol{\mathsf{S}}\right\rrbracket \boldsymbol{\mathsf{H}}\right),\label{eq:version1}
\end{align}
where the first part is simply the likelihood in the absence of deterministic signals (Eq. (\ref{eq:marginalized likelihood})), and the sky location of the burst $(\theta,\phi)$ only appears in the two inner products $\left\llbracket \boldsymbol{r}|\boldsymbol{\mathsf{S}}\right\rrbracket $
and $\left\llbracket \boldsymbol{\mathsf{S}}|\boldsymbol{\mathsf{S}}\right\rrbracket$. Since we do not have any information about the burst waveform, elements in $\mathsf{\boldsymbol{\mathsf{H}}}$ can in principle take any real values. A reasonable prior on $\mathsf{\boldsymbol{\mathsf{H}}}$ is the Gaussian distribution:
\begin{equation}
\pi(\mathsf{\boldsymbol{\mathsf{H}}}|q)=\mathcal{N}(\mathsf{\boldsymbol{\mathsf{H}}}|\boldsymbol{Q}),
\end{equation}
where the covariance matrix is defined to have the form $\boldsymbol{Q}=10^{2q}\boldsymbol{I}$, with $\boldsymbol{I}$ being a $2n_{\tiny{H}}\times2n_{\tiny{H}}$ identity matrix.\footnote{Waveform parameters from a certain physical process should be correlated in some way, so the covariance matrix should have off-diagonal entries. However, considering that we do not have any a priori information about the signal, and that the number of pieces $n_{\tiny H}$ is much smaller than the number of TOAs, a diagonal $\boldsymbol{Q}$ may not be a bad assumption. } The entries in $\boldsymbol{Q}$ have unit $\text{[s}^{2}]$. $\mathsf{\boldsymbol{\mathsf{H}}}$ plays a similar role as $\boldsymbol{b}$ in Eq. (\ref{eq:bB}) and $q$ is similar to $\boldsymbol{\eta}$ (such as $A$ and $\gamma$). Note that elements in $\mathsf{\boldsymbol{\mathsf{H}}}$ are expected to be not much larger than $10^{-6}\ \text{s}$, which is the order of magnitude of the residuals $\boldsymbol{r}$. Hence the hyperparameter $q$ should have a value comparable to or smaller than $-6$.

With the likelihood $\mathcal{L}(\boldsymbol{r}|\boldsymbol{\eta},\theta,\phi,\boldsymbol{\mathsf{H}})$ and the prior $\pi(\mathsf{\boldsymbol{\mathsf{H}}}|q)$ (in addition to the trivial priors on other parameters), we are able to perform a Bayesian analysis in search of the posterior distributions for all model parameters. However, before searching for a burst signal with a PTA data set, the first question one may ask is ``is there a burst in the data?'' rather than ``what is the waveform?''. If we are only interested in whether a burst exists, note that since both $\mathcal{L}(\boldsymbol{r}|\boldsymbol{\eta},\theta,\phi,\boldsymbol{\mathsf{H}})$ and $\pi(\mathsf{\boldsymbol{\mathsf{H}}}|q)$ are Gaussian functions for $\mathsf{\boldsymbol{\mathsf{H}}}$, we can integrate out the waveform analytically and obtain the marginalized likelihood:
\begin{widetext}
\begin{align}
\mathcal{L}(\boldsymbol{r}|\boldsymbol{\eta},\theta,\phi,q) & =\int\mathcal{L}(\boldsymbol{r}|\boldsymbol{\eta},\theta,\phi,\boldsymbol{\mathsf{H}})\pi(\mathsf{\boldsymbol{\mathsf{H}}}|q)\text{d}\mathsf{\boldsymbol{\mathsf{H}}}\nonumber\\
 & =\int\mathcal{N}(\boldsymbol{r}|\boldsymbol{C})\frac{\exp\left[\left\llbracket \boldsymbol{r}|\boldsymbol{\mathsf{S}}\right\rrbracket \boldsymbol{\mathsf{H}}-\frac{1}{2}\boldsymbol{\mathsf{H}}^{\top}\left(\left\llbracket \boldsymbol{\mathsf{S}}|\boldsymbol{\mathsf{S}}\right\rrbracket +\boldsymbol{Q}^{-1}\right)\boldsymbol{\mathsf{H}}\right]}{\sqrt{\det\left(2\pi\boldsymbol{Q}\right)}}\text{d}\mathsf{\boldsymbol{\mathsf{H}}}\nonumber\\
 & =\mathcal{N}(\boldsymbol{r}|\boldsymbol{C})\frac{\exp\left(\frac{1}{2}\left\llbracket \boldsymbol{r}|\boldsymbol{\mathsf{S}}\right\rrbracket \boldsymbol{\Sigma}^{-1}\left\llbracket \boldsymbol{\mathsf{S}}|\boldsymbol{r}\right\rrbracket \right)}{\sqrt{\det\left(\boldsymbol{Q}\boldsymbol{\Sigma}\right)}},\label{eq:version2}
\end{align} 
\end{widetext}
where we have defined $\boldsymbol{\Sigma}\equiv\left\llbracket\boldsymbol{\mathsf{S}}|\boldsymbol{\mathsf{S}}\right\rrbracket +\boldsymbol{Q}^{-1}$ in the last step.\footnote{Using the Woodbury identity, $\mathcal{L}(\boldsymbol{r}|\boldsymbol{\eta},\theta,\phi,q)$  can be written in a more compact form: $\mathcal{L}(\boldsymbol{r}|\boldsymbol{\eta},\theta,\phi,q)=\mathcal{N}(\boldsymbol{r}|\boldsymbol{C}_{q})$, where $\boldsymbol{C}_{q}\equiv\boldsymbol{C}+\boldsymbol{\mathsf{S}}\boldsymbol{Q}\boldsymbol{\mathsf{S}}^{\top}$. This is of the same form as the likelihood in Eq. (\ref{eq:marginalized likelihood}); the difference is the extra term $\boldsymbol{\mathsf{S}}\boldsymbol{Q}\boldsymbol{\mathsf{S}}^{\top}$.} Therefore, without reconstructing the burst waveform, we are able to search for the noise parameters in $\boldsymbol{C}$ and the sky location of the burst in $\boldsymbol{\mathsf{S}}$ (if there was indeed a burst). The hyperparameter $q$ in $\boldsymbol{Q}$ controls the prior of $\mathsf{\boldsymbol{\mathsf{H}}}$. If $q$ is fixed to be a small number, e.g., $q=-9$, only tiny values ($\lesssim10^{-9}$ in the unit of {[}s{]}) would be assigned to $\mathsf{\boldsymbol{\mathsf{H}}}$, and hence the search is effectively equivalent to searching in a noise-only model. Therefore, the ratio of $q$'s prior and posterior at small $q$ gives the Savage-Dickey density ratio \cite{dickey1971weighted}, which is equivalent to the Bayes factor comparing our model to the model without deterministic signals. If $q$'s marginal posterior does not have support near the lower bound of its prior, our model is strongly preferred over the null model.

\subsection*{Waveform reconstruction\label{subsec:Waveform-reconstruction}}

If our model is favored over the noise-only model, one would then be interested in what the burst waveform looks like. In order to reconstruct the waveform, a straightforward way is to go back to the likelihood given by Eq. (\ref{eq:version1}) and perform a Bayesian analysis over all parameters. Depending on how many pieces into which we divide the observation period, the number of parameters in $\mathsf{\boldsymbol{\mathsf{H}}}$ could be large, which makes the search computationally expensive. However, it turns out the cost can be reduced significantly if we exploit the samples drawn from the posterior based on Eq. (\ref{eq:version2}) and search for the waveform parameters one by one.

Note again that the part containing $\mathsf{\boldsymbol{\mathsf{H}}}$ in Eq. (\ref{eq:version1}) is (part of) a multivariate Gaussian distribution, which has the property that if some of the variables are integrated out, the rest also obey a multivariate Gaussian distribution. The means would be the corresponding means of the original distribution, and the covariance matrix would be the corresponding submatrix of the original one. For example, let $\mathcal{N}(\boldsymbol{x}|\boldsymbol{\mu},\boldsymbol{D})$ denote a Gaussian distribution with mean $\boldsymbol{\mu}$ and covariance $\boldsymbol{D}$. If all variables except for $x_{i}$ are integrated out, the marginalized distribution of $x_{i}$ is
\begin{align}
p(x_{i}) & =\int\mathcal{N}(\boldsymbol{x}|\boldsymbol{\mu},\boldsymbol{D})\text{d}x_{1}\text{d}x_{2}...\text{d}x_{i-1}\text{d}x_{i+1}...\nonumber\\
&=\mathcal{N}(x_{i}|\mu_{i},{D}_{ii}),
\end{align}
where $\mu_{i}$ is the $i$th element of vector $\boldsymbol{\mu}$ and ${D}_{ii}$ represents the $i$th diagonal element in matrix $\boldsymbol{D}$. In our context, if, for example, we integrate out the last $2n_{\tiny{H}}-1$ elements in $\mathsf{\boldsymbol{\mathsf{H}}}=\begin{pmatrix}\mathsf{H}_{1}^{+} & \mathsf{H}_{2}^{+} & \cdots & \mathsf{H}_{n_{\tiny{H}}}^{+} & \mathsf{H}_{1}^{\times} & \mathsf{H}_{2}^{\times} & \cdots & \mathsf{H}_{n_{\tiny{H}}}^{\times}\end{pmatrix}^{\top}$, the marginalized likelihood becomes
\begin{widetext}
\begin{align}
\mathcal{L}(\boldsymbol{r}|\boldsymbol{\eta},\theta,\phi,q,\mathsf{H}_{1}^{+}) & =\int\mathcal{L}(\boldsymbol{r}|\boldsymbol{\eta},\theta,\phi,\mathsf{\boldsymbol{\mathsf{H}}})\pi(\mathsf{\boldsymbol{\mathsf{H}}}|q)\text{d}\mathsf{H}_{2}^{+}\text{d}\mathsf{H}_{3}^{+}...\text{d}\mathsf{H}_{n_{\tiny{H}}}^{\times}\nonumber\\
 & =\mathcal{N}(\boldsymbol{r}|\boldsymbol{C})\frac{\exp\left(\frac{1}{2}\left\llbracket \boldsymbol{r}|\boldsymbol{\mathsf{S}}\right\rrbracket \boldsymbol{\Sigma}^{-1}\left\llbracket \boldsymbol{\mathsf{S}}|\boldsymbol{r}\right\rrbracket \right)}{\sqrt{\det\left(\boldsymbol{Q}\boldsymbol{\Sigma}\right)}}\int\mathcal{N}\left(\mathsf{\boldsymbol{\mathsf{H}}}|\boldsymbol{\Sigma}^{-1}\left\llbracket \boldsymbol{\mathsf{S}}|\boldsymbol{r}\right\rrbracket ,\boldsymbol{\Sigma}\right)\text{d}\mathsf{H}_{2}^{+}\text{d}\mathsf{H}_{3}^{+}...\text{d}\mathsf{H}_{n_{\tiny{H}}}^{\times}\nonumber\\
 & =\mathcal{L}(\boldsymbol{r}|\boldsymbol{\eta},\theta,\phi,q)\mathcal{N}\left(\mathsf{H}_{1}^{+}|\left(\boldsymbol{\Sigma}^{-1}\left\llbracket \boldsymbol{\mathsf{S}}|\boldsymbol{r}\right\rrbracket \right)_{1},{\Sigma}_{11}\right)\label{eq:version3}
\end{align}    
\end{widetext}
where in the last step we have used Eq. (\ref{eq:version2}). A similar expression applies to all other parameters in $\mathsf{\boldsymbol{\mathsf{H}}}$. This allows us to perform the search of $\mathsf{\boldsymbol{\mathsf{H}}}$ one element after another. 

This can be achieved as follows. In performing a Bayesian analysis based on the likelihood given by Eq. (\ref{eq:version2}), we have obtained the posterior $p(\boldsymbol{\eta},\theta,\phi,q|\boldsymbol{r})\propto \mathcal{L}(\boldsymbol{r}|\boldsymbol{\eta},\theta,\phi,q)$ from, e.g., MCMC sampling. To find the marginal posterior distribution of, say, $\mathsf{H}_{1}^{+}$, we need to integrate out parameters $\boldsymbol{\eta},\theta,\phi$ and $q$ in Eq. (\ref{eq:version3}), which contains the distribution $\propto \mathcal{L}(\boldsymbol{r}|\boldsymbol{\eta},\theta,\phi,q)$. To find the posterior density at $\mathsf{H}_{1}^{+}=x$, we simply need to take the sum of $\mathcal{N}\left(x|\left(\boldsymbol{\Sigma}^{-1}\left\llbracket \boldsymbol{\mathsf{S}}|\boldsymbol{r}\right\rrbracket \right)_{1},{\Sigma}_{11}\right)$ over all the samples. In other words,
\begin{equation}
p(\mathsf{H}_{1}^{+}=x)\propto \sum_{j}\mathcal{N}\left(x|\left(\boldsymbol{\Sigma}^{-1}\left\llbracket \boldsymbol{\mathsf{S}}|\boldsymbol{r}\right\rrbracket \right)_{1}^{(j)},{\Sigma}_{11}^{(j)}\right),\label{eq:MCMC_A}
\end{equation}
where the superscript ``$(j)$'' represents the $j$th sample in the chain. In principle, $x$ can take any value, but in the presence of a burst, we would expect that $\mathsf{H}_{1}^{+}$ only has support near $\left(\boldsymbol{\Sigma}_{*}^{-1}\left\llbracket \boldsymbol{\mathsf{S}}_{*}|\boldsymbol{r}\right\rrbracket \right)_{1}$
within the range $\sim{\Sigma}_{*11}$, where ``{$_*$}'' denotes the maximum-a-posteriori value obtained from  $p(\boldsymbol{\eta},\theta,\phi,q|\boldsymbol{r})$. Therefore, we can simply make a grid near this region and evaluate the posterior of $\mathsf{H}_{1}^{+}$. This process can be repeated for other parameters in $\mathsf{\boldsymbol{\mathsf{H}}}$.

In conclusion, there are two practical ``versions'' of likelihood in our model. If one wants to know whether a burst signal exists, we simply need to use Eq. (\ref{eq:version2}) that efficiently gives the marginal posterior for $q$. If a burst signal is present, the sky location $(\theta,\phi)$ can also be tracked down from this analysis. If we are interested in how the burst looks like, we could then use Eqs. (\ref{eq:version3}) and (\ref{eq:MCMC_A}) to find the posterior distributions of $\mathsf{\boldsymbol{\mathsf{H}}}$ without a further Bayesian search. It will be shown in the next section how these two likelihoods are applied in simulated data sets. 

\begin{table}
\begin{tabular}{|c|c|c|}
\hline 
Parameter & Prior range (uniform) & True value\tabularnewline
\hline 
$\log_{10}A$ & $[-18,-13]$ & -14.398\tabularnewline
$\gamma$ & $[0,7]$ & 13/3\tabularnewline
$\cos\theta$ & $[-1,1]$ & 0.5\tabularnewline
$\phi$ & $[0,2\pi]$ & 3\tabularnewline
$q$ & $[-9,-5]$ & -\tabularnewline
\hline 
\end{tabular}

\caption{\label{tab:Prior}Prior distributions for $\log_{10}A,\gamma,\cos\theta,\phi$
and $q$  and the true values in the simulated data sets.}
\end{table}

\section{Analyses of simulated data sets \label{simulated data}}

In this section, we test our model by analyzing three simulated data sets, each consisting of 20 pulsars. Each pulsar has been observed for 10 years every 30 days. Since the time scale {and the shape} of the burst signal are unknown, the ideal number of grid points that divide the TOAs and the ideal grid spacing cannot be determined beforehand and so should be regarded as parameters. However, fixing the grid can significantly reduce the computational cost.  In this work, we divide the PTA observation period into 20 even pieces ($n_{\tiny{H}}=21$), leaving the effect of varying $n_{\tiny{H}}$ and the grid spacing to future work (see discussion in Sec. \ref{discussion}).

For simplicity, all simulated residuals have the same constant white noise level of $0.5\ \mu$s and no intrinsic pulsar red noise. In addition, an SGWB is injected with a power spectrum given by Eq. (\ref{eq:power}), with $A=4\times10^{-15}$ ($\log_{10}A\approx-14.398$) and $\gamma=13/3$. Although the injected SGWB induces correlated red noise among pulsars, in the following analyses we treat the background as an uncorrelated common red noise process. This greatly reduces the computational cost because the noise matrix $\boldsymbol{C}$ is now block-diagonal, which allows  $\boldsymbol{C}^{-1}$ to be computed block by block {(or pulsar by pulsar)}. Consequently, the inner products $\left\llbracket \boldsymbol{r}|\boldsymbol{\mathsf{S}}\right\rrbracket $
and $\left\llbracket \boldsymbol{\mathsf{S}}|\boldsymbol{\mathsf{S}}\right\rrbracket$ can be obtained rather efficiently. {Ignoring pulsar correlations in $\boldsymbol{C}$ could bias the burst search by overestimating the significance of the potential signal, but since the correlations should have a smaller effect compared to the common spectrum, using a block-diagonal $\boldsymbol{C}$ is not expected to significantly affect the results in the simple scenarios we are considering in this work. }

Following Ref. \cite{Finn:2010ph}, a burst signal from the parabolic encounter of two SMBHs is injected. {The waveform in Ref. \cite{Finn:2010ph} was obtained from the quadrupole formula applied to a parabolic Kepler orbit. It serves as a first approximation of the encounter event and as an exemplar for the purpose of  demonstrating how our model works.} The event under consideration is set to occur at sky location $(\cos\theta,\phi)=(0.5,3)$. The two black holes have the same mass $10^{9}M_{\odot}$, and the impact parameter is $2\times10^{11}M_{\odot}$ {(where we have set $G=c=1$)}. The injected signal is also set to sit in the middle of the observation period.

Similar to Ref. \cite{Finn:2010ph}, we shall test our model with data sets of different signal-to-noise ratios (SNRs, defined by $\text{SNR}=\sqrt{\left\llbracket \boldsymbol{h}|\boldsymbol{h}\right\rrbracket }$). In the first two data sets,  the encounter event occurs at different distances from us: 20 Mpc (strong signal, $\text{SNR}\approx14.7$) and 45 Mpc (weak signal, $\text{SNR}\approx6.5$). For each case, we first perform an analysis based on the noise-only model, with the likelihood given by Eq. (\ref{eq:marginalized likelihood}), which only contains two parameters: $A$ and $\gamma$; then we search for the existence and sky location of the burst using the likelihood  given by Eq. (\ref{eq:version2}), which contains five parameters: $A,\gamma,\theta,\phi$ and $q$; lastly, we reconstruct the waveform according to Eq. (\ref{eq:MCMC_A}), and compare the results with the injected signals. In addition, we also test our model with a data set that contains no burst signals. The priors on $\log_{10}A,\gamma,\cos\theta,\phi$ and $q$ are all set as uniform distributions, with bounds shown in Table \ref{tab:Prior}. In what follows, the Bayesian inferences are achieved by Nestle,\footnote{https://github.com/kbarbary/nestle} {a Python implementation of the nested sampling algorithm \cite{skilling2004nested,sivia2006data,Shaw:2007jj,Mukherjee:2005wg,Feroz:2008xx} aiming at comparing models and generating samples from posterior distributions.}
\begin{figure}
\centering
\includegraphics[scale=0.4]{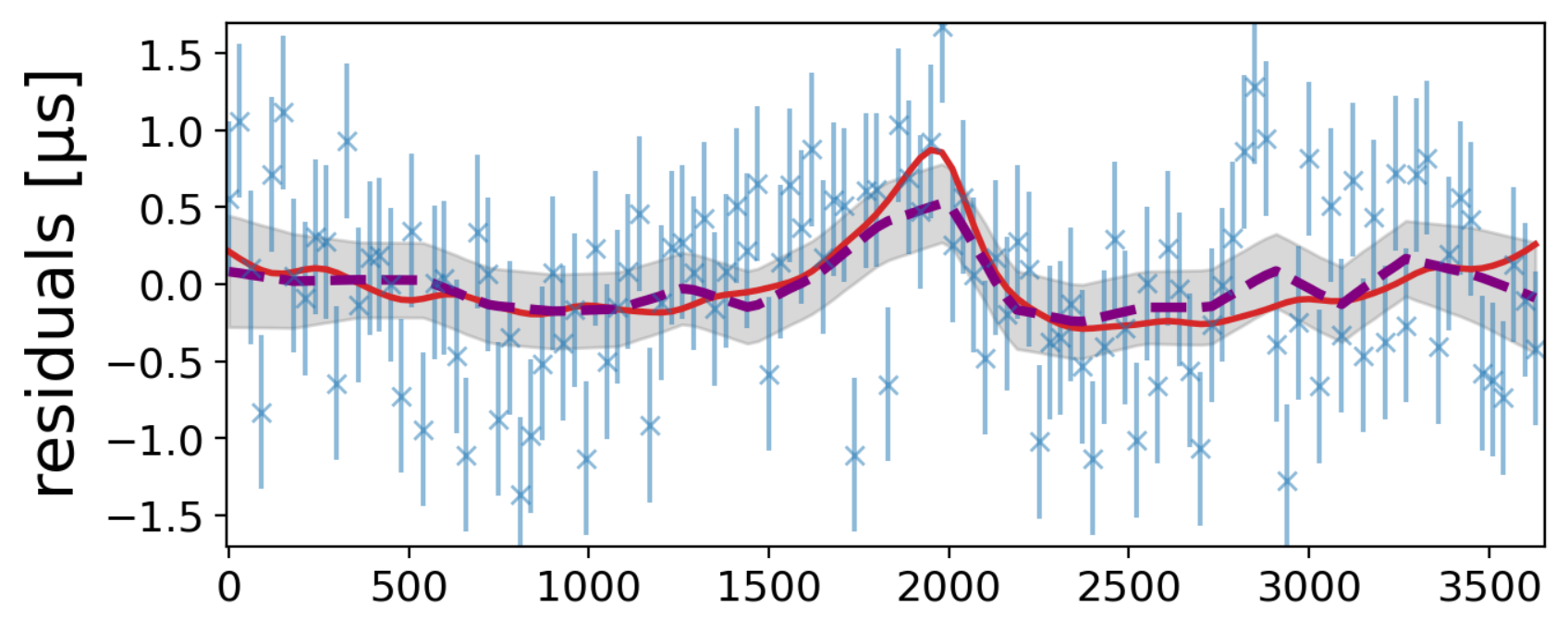}
\includegraphics[scale=0.4]{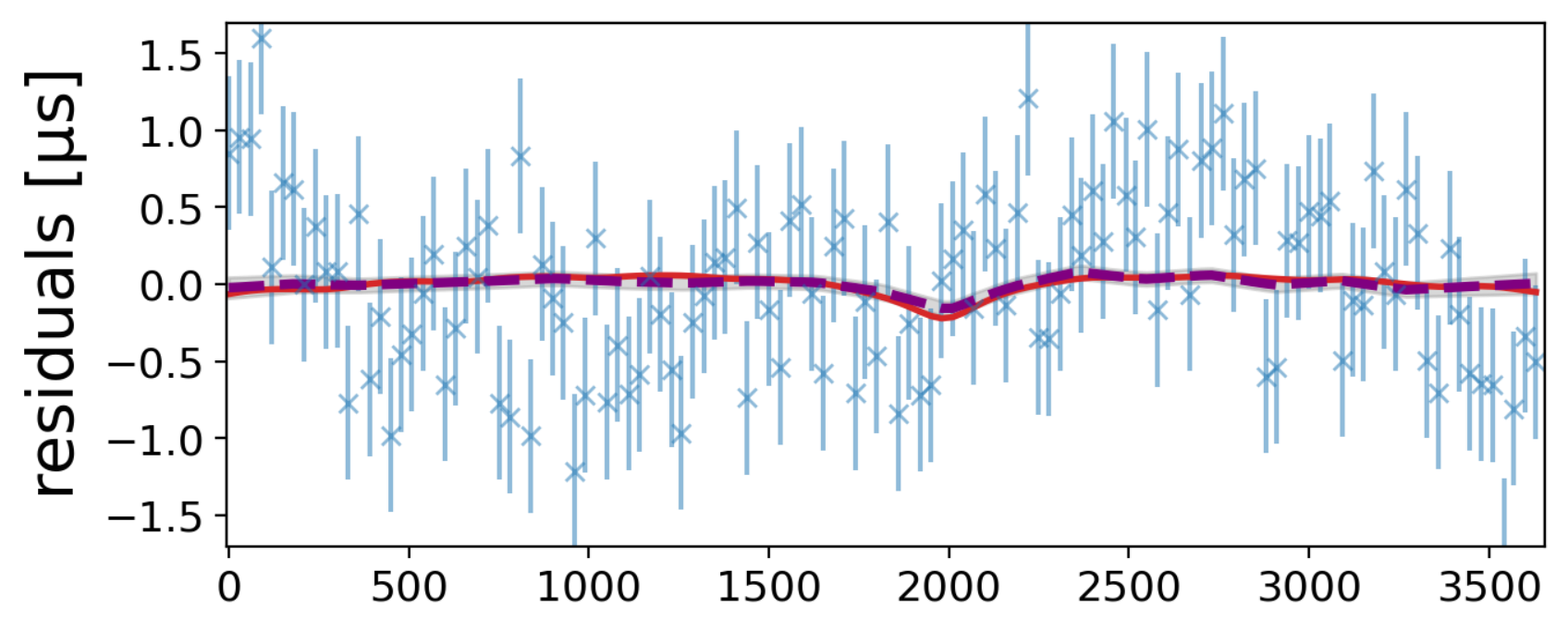}
\includegraphics[scale=0.4]{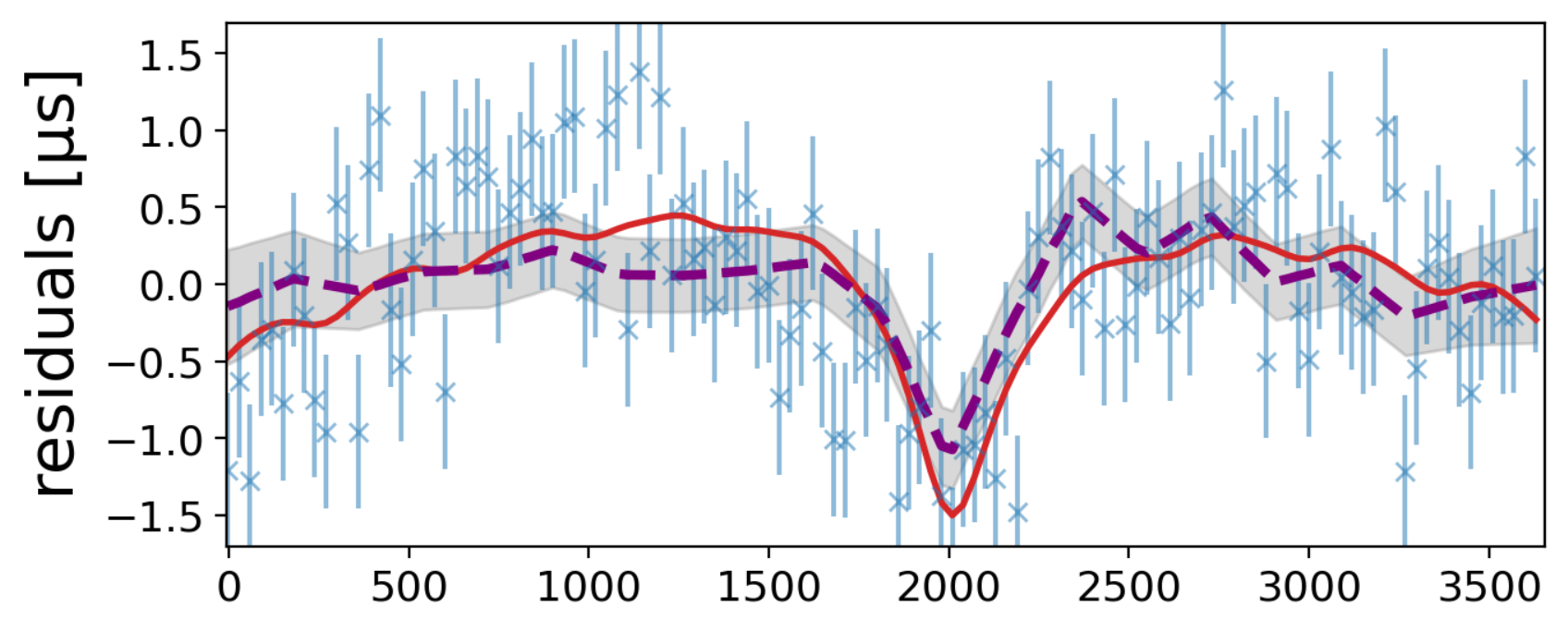}
\includegraphics[scale=0.4]{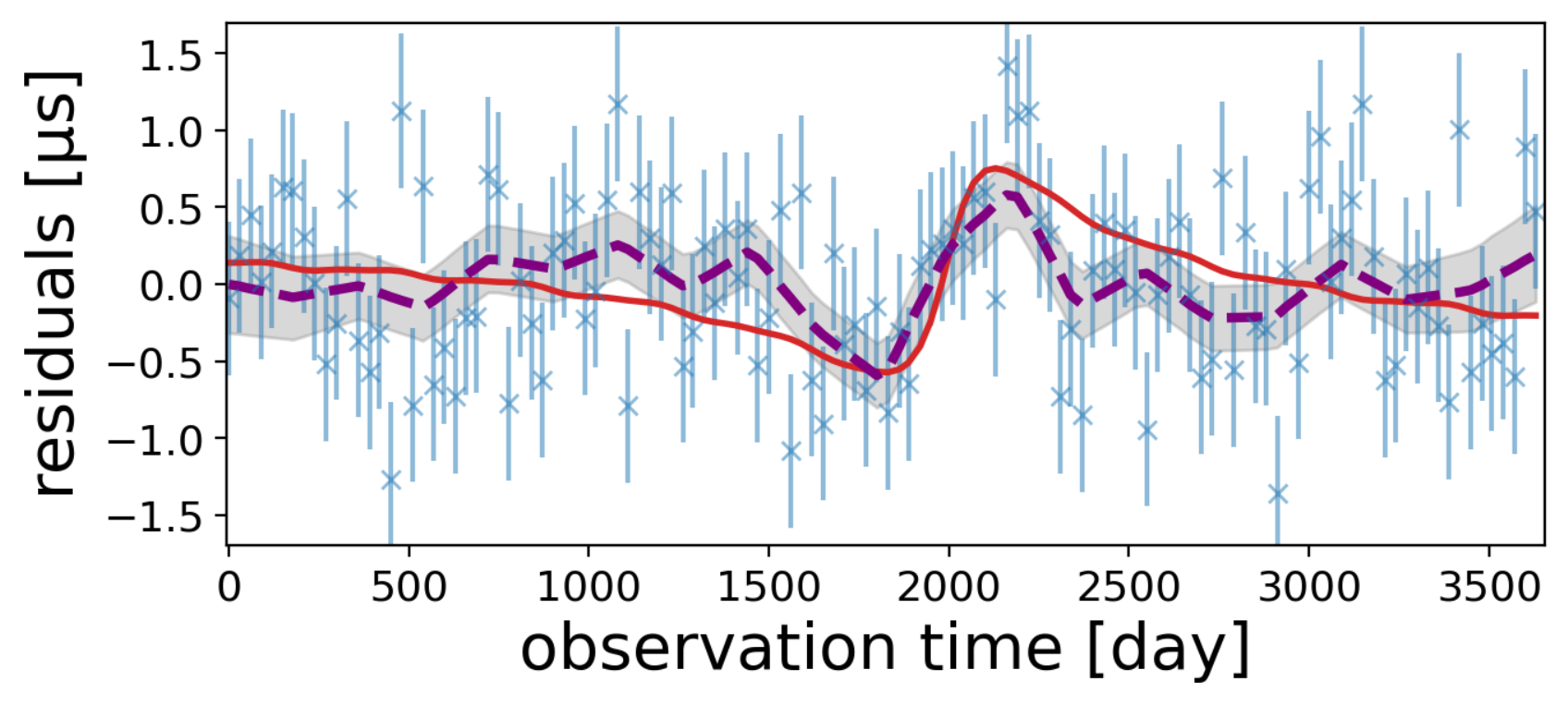}
\caption{\label{fig:PTA-1}Post-fit residuals (blue data points with error bars) and perturbations from the injected strong burst (solid red curves) for 4 of the 20 pulsars. {The reconstructed burst-induced residuals are shown as dashed, purple curves (with 90\% confidence intervals), where the sky location of the burst is taken as the  posterior medians for $\cos\theta$  and $\phi$.}}
\end{figure}

\begin{figure}
\centering
\includegraphics[scale=0.5]{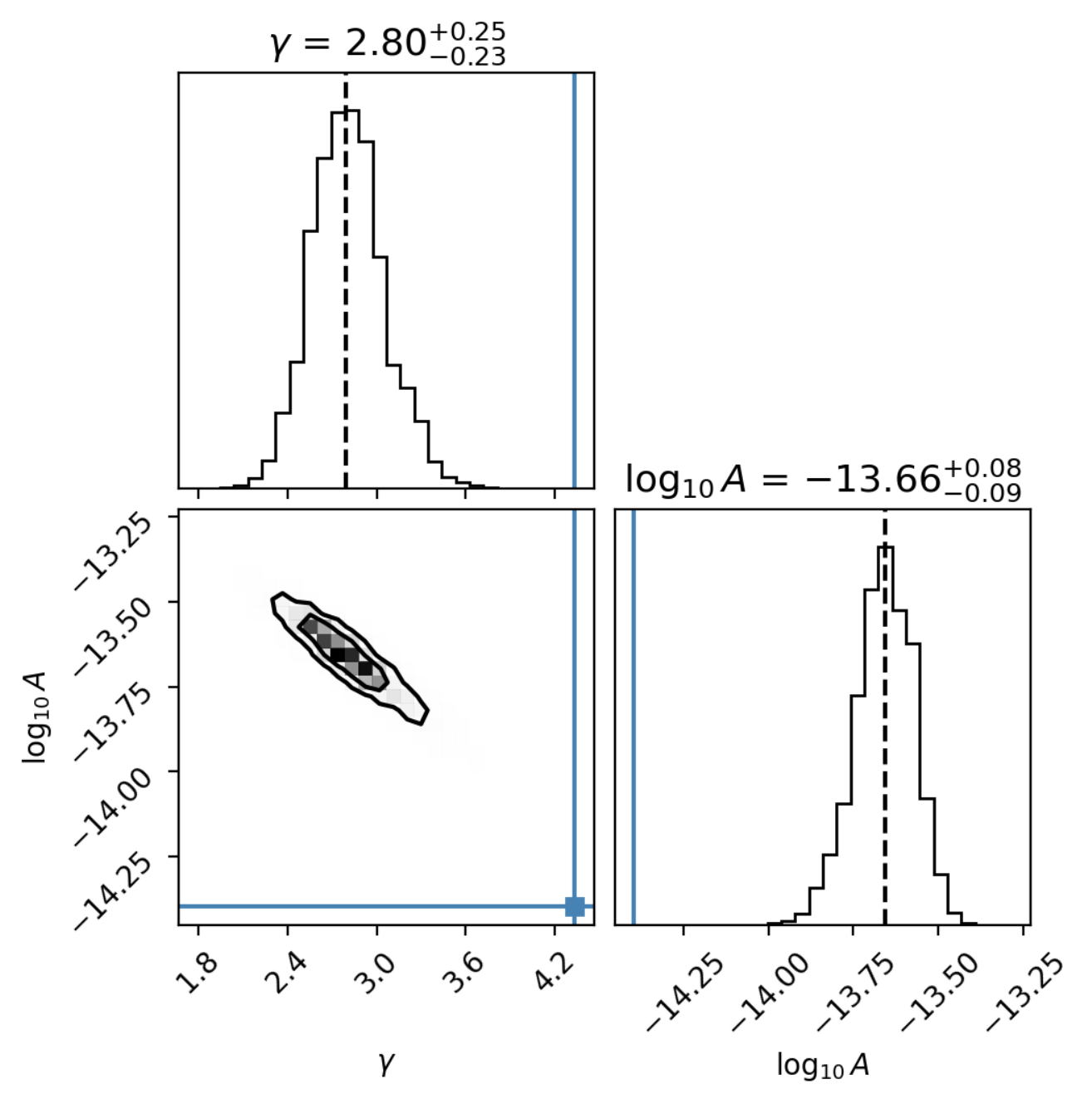}
\caption{\label{fig:null1}Posterior distributions of the SGWB parameters $A$ and $\gamma$ in the noise-only model when a strong burst signal is present. The dashed lines represent the median values. The distributions are obviously incompatible with the true values in the simulated data set ($\log_{10}A\approx-14.398$ and $\gamma\approx4.33$) represented by blue lines}
\end{figure}

\begin{figure*}
\centering
\includegraphics[scale=0.5]{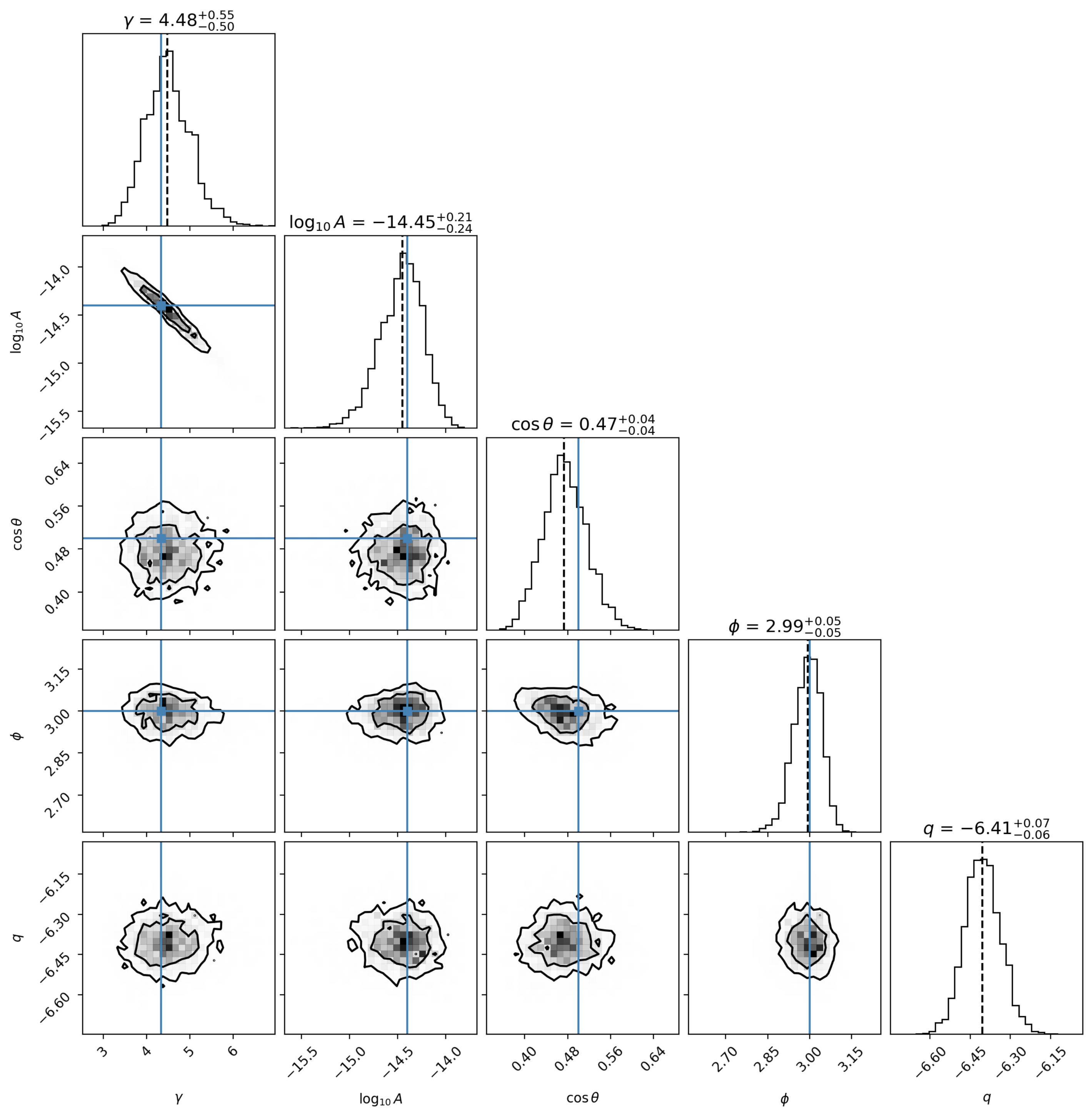}

\caption{\label{fig:model1}Posterior distributions of parameters $\log_{10}A,\gamma,\cos\theta,\phi$
and $q$ in our model  when a strong burst signal is present. The dashed lines represent the median values, and the blue solid lines denote the true values in the simulated data set. {The red noise parameters are recovered and the location of the burst is detected. The existence of the burst has significant evidence since the hyperparameter $q$ has no samples near the lower bound of its prior ($q_{\rm min}=-9$).}}
\end{figure*}

\subsection{Strong signal ($\text{SNR}\approx14.7$)}

We first consider a case where a strong signal is injected into the background. The source is placed at a distance of 20 Mpc. In Fig. \ref{fig:PTA-1} we show the post-fit timing residuals\footnote{{Here ``post-fit'' means the contribution in residuals fitted by the timing model, such as the quadratic components, has been removed.}} and the contribution from the injected burst for 4 of the 20 pulsars. We first analyze the data set with the noise-only model, where the residuals are assumed to be generated by white noise and an SGWB. With the white noise parameters fixed, there are only two free parameters: the amplitude of the SGWB spectrum $A$ and the spectral index $\gamma$. Using the likelihood  given by Eq. (\ref{eq:marginalized likelihood}), we obtain the posterior distributions shown in Fig. \ref{fig:null1}. The median values ($\log_{10}A\approx-13.66$ and $\gamma\approx2.79$) obviously deviate from the true values ($\log_{10}A\approx-14.398$ and $\gamma\approx4.33$). In the presence of the strong signal, the noise-only model is not able to recover the SGWB faithfully. {Since the noise-only model treats the burst signal as part of the background, the ``detected'' magnitude of the SGWB spectrum is larger than the true value.}

We then perform a Bayesian analysis with the likelihood given by Eq. (\ref{eq:version2}), which contains three additional parameters: $\theta,\phi$ and $q$. The posterior distributions are shown in Fig. \ref{fig:model1}. We can see that the SGWB parameters are now captured by our model, with median values and true values being rather close to each other.\footnote{How well the median and true values agree depends on the realization of simulated data. With the background parameters (white noise level, $A$ and $\gamma$) fixed, different realizations of the stochastic feature (i.e., different sets of random number generators) can lead to statistical errors.}  The hyperparameter $q$ is only sampled near $-6.4$, which indicates a large Savage-Dickey ratio. In fact, the Bayes factor for our model to the noise-only model ({based on the Nestle results}) is $\sim10^{10}$, corresponding to overwhelming evidence for the presence of a signal. From the marginal posterior distributions of $\cos\theta$ and $\phi$, the injected burst is also perfectly localized on the sky map by our model. 

\begin{figure}
\centering
\includegraphics[scale=0.45]{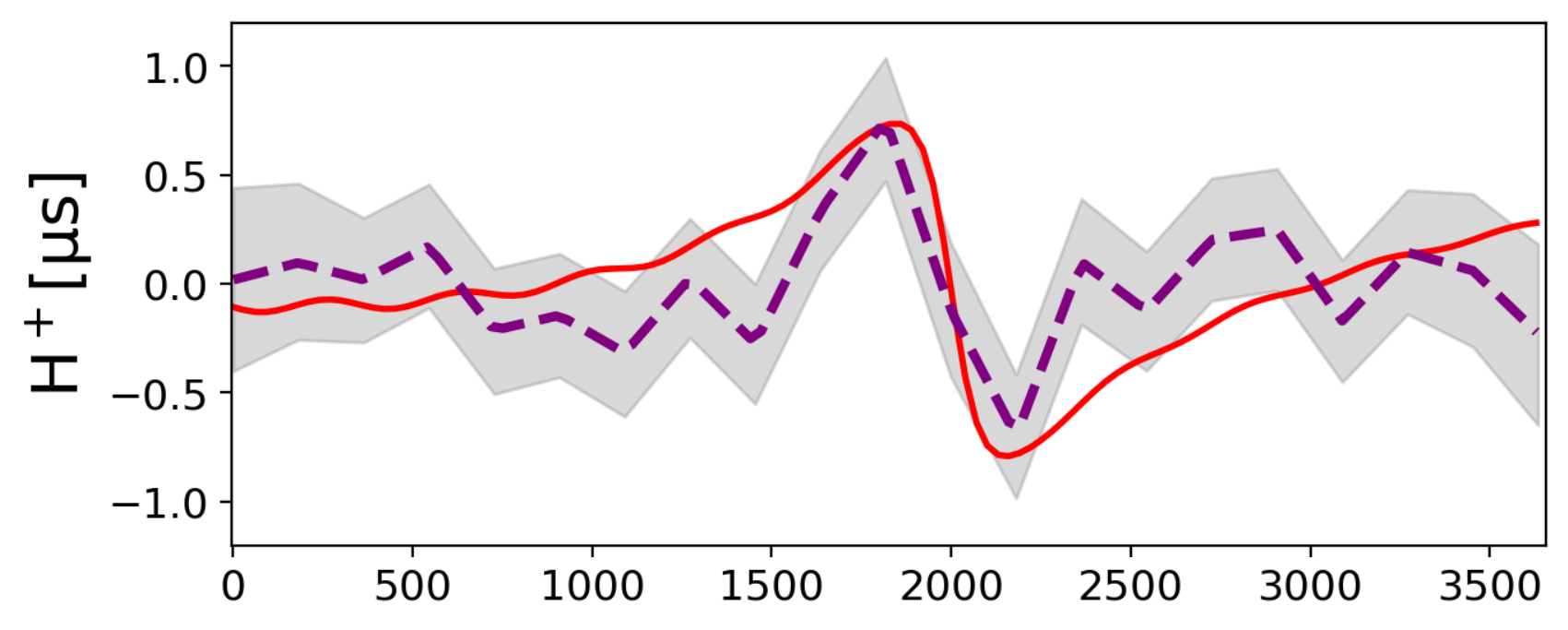}
\includegraphics[scale=0.45]{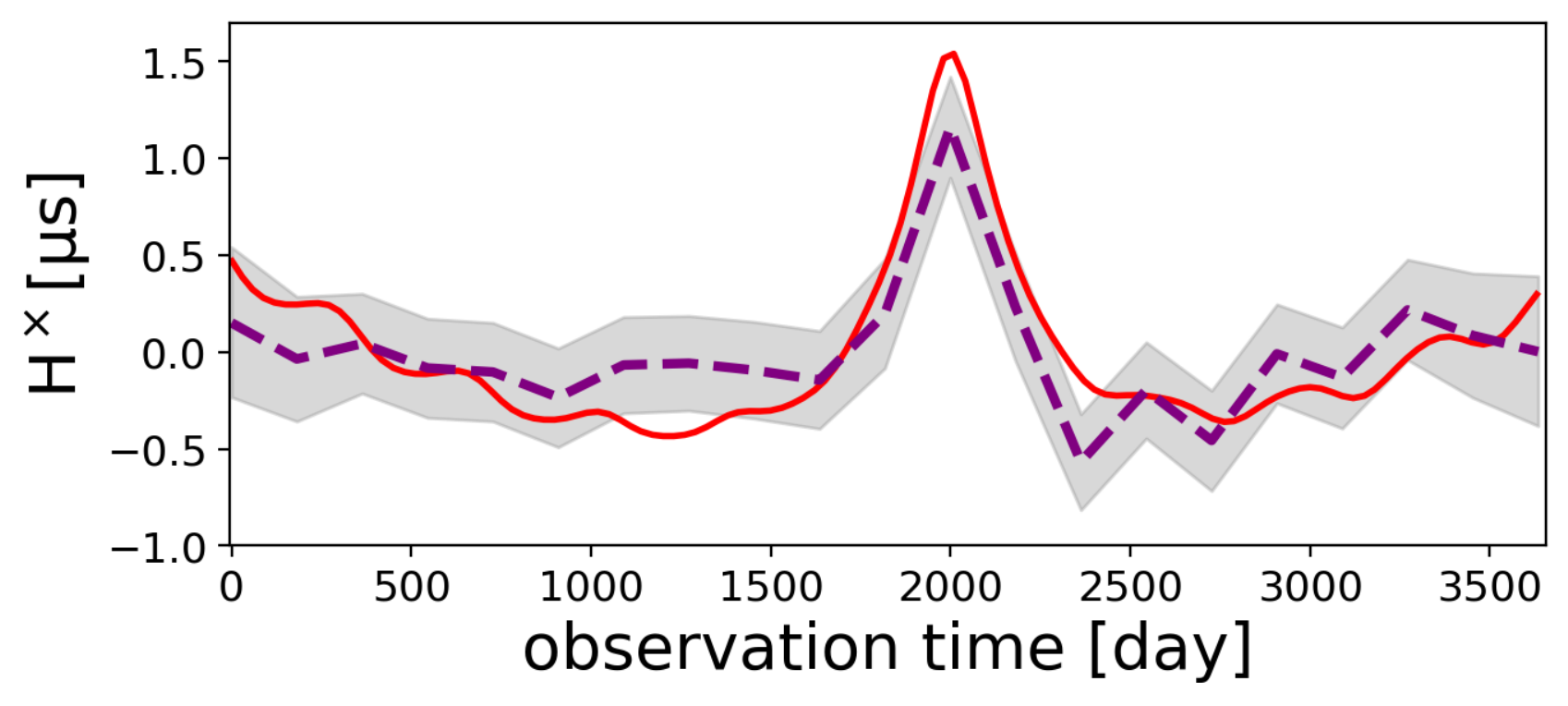}
\caption{\label{fig:waveform_reconstruction}Injected $H^{+}(t)$ and $H^{\times}(t)$ (post-fit) for a strong burst signal (red curves), the reconstructed piecewise linear functions $\mathsf{H}_{\mu}^{+}$ and $\mathsf{H}_{\mu}^{\times}$ (dashed, purple curves, median values) and 90\% confidence intervals (shaded region), {where the sky location of the burst is taken as the  posterior medians for $\cos\theta$  and $\phi$.}}
\end{figure}

Following the method described in Subsec. \ref{subsec:Waveform-reconstruction}, we then use the samples from the last paragraph to infer the burst waveform. The best-fit piecewise straight lines describing $H^{+}$ and $H^{\times}$ are shown in Fig. \ref{fig:waveform_reconstruction} by dashed, purple curves with 90\% confidence intervals. The red curves are the post-fit injected $H^{+}$ and $H^{\times}$. {We also show in Fig. \ref{fig:PTA-1}  the reconstructed burst-induced residuals for 4 pulsars. }

\begin{figure}
\centering
\vspace{0.1in}
\includegraphics[scale=0.4]{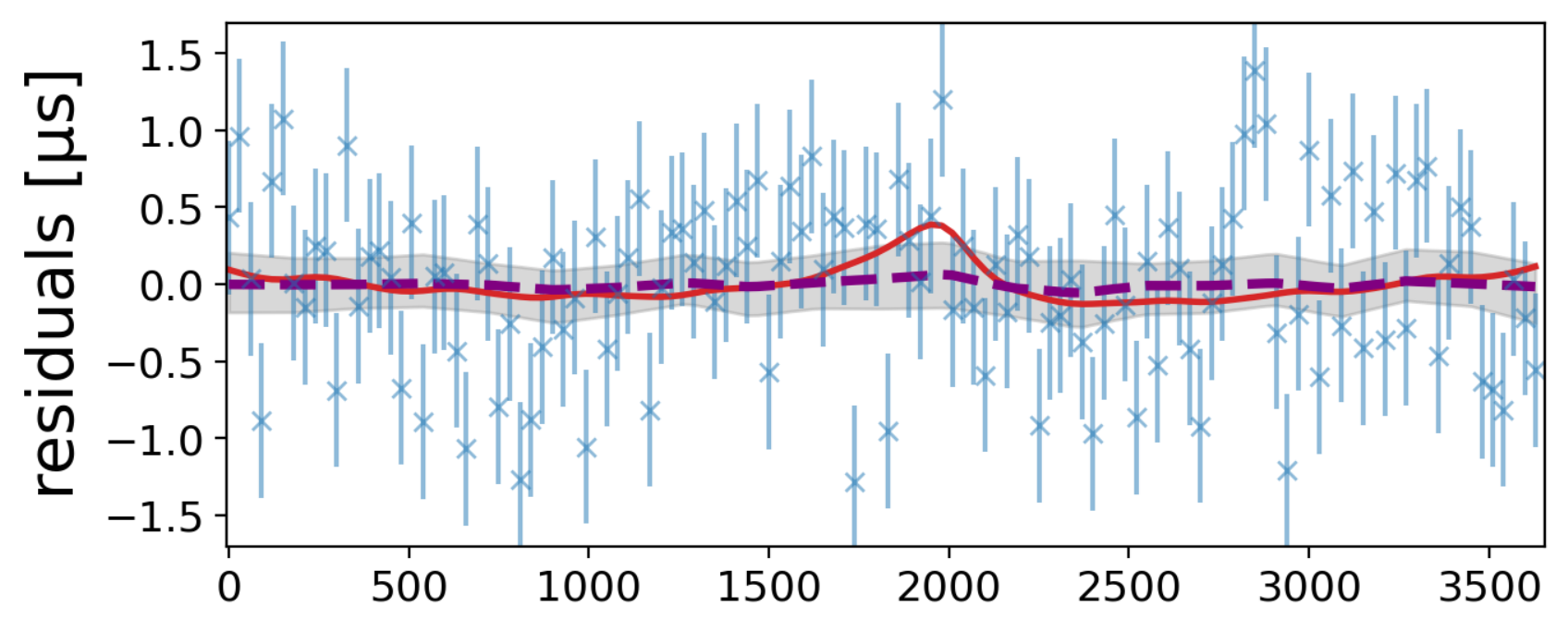}
\includegraphics[scale=0.4]{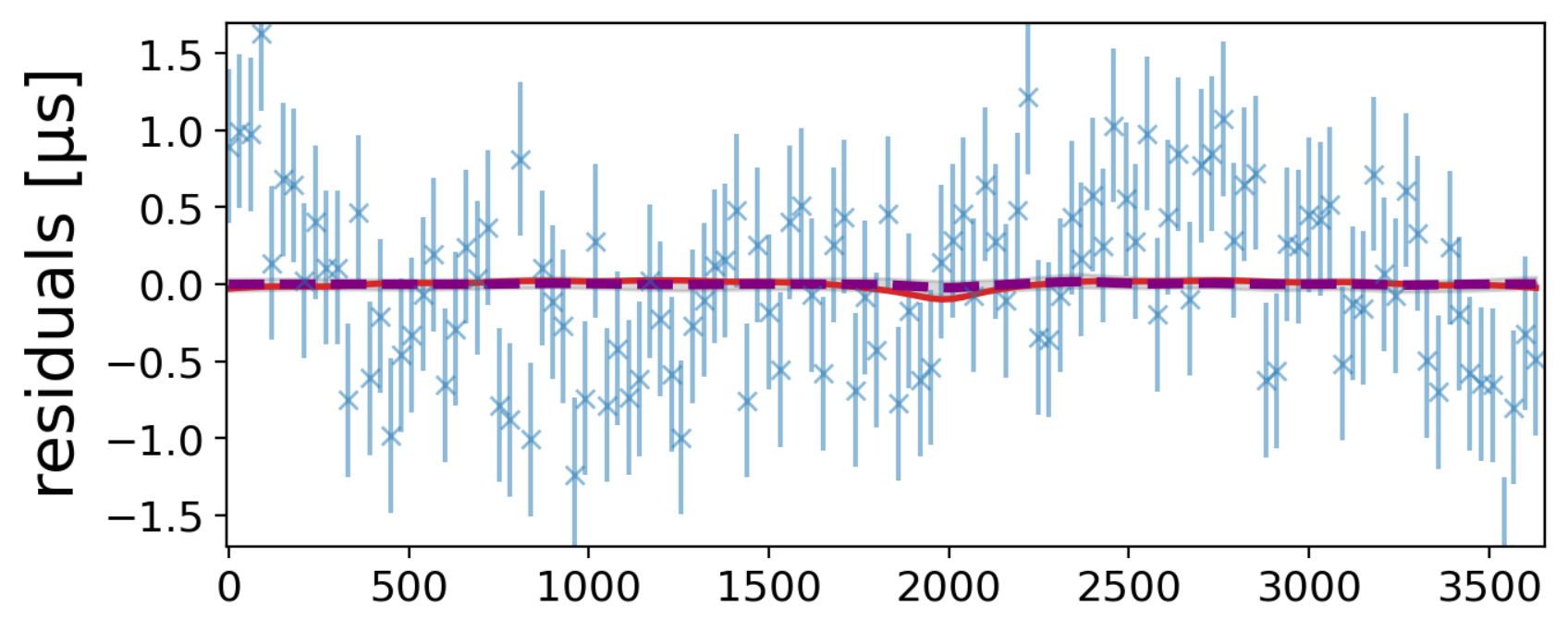}
\includegraphics[scale=0.4]{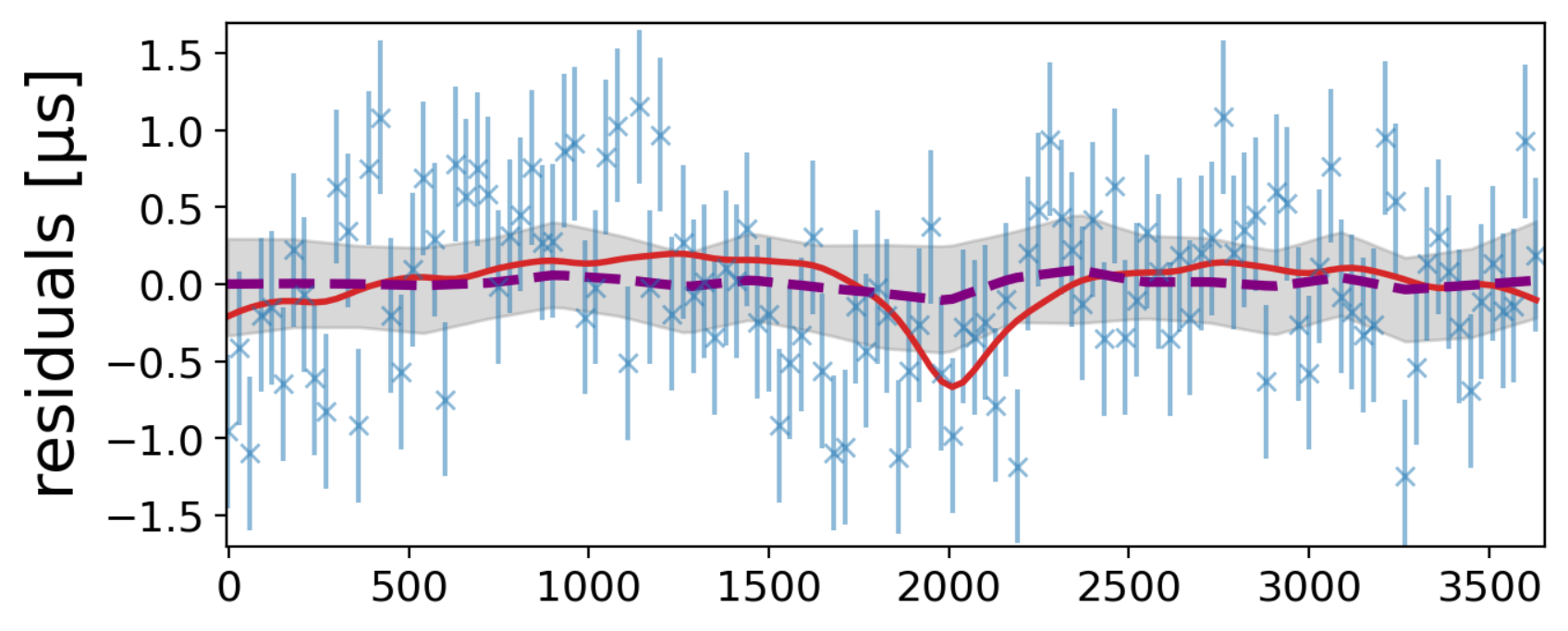}
\includegraphics[scale=0.4]{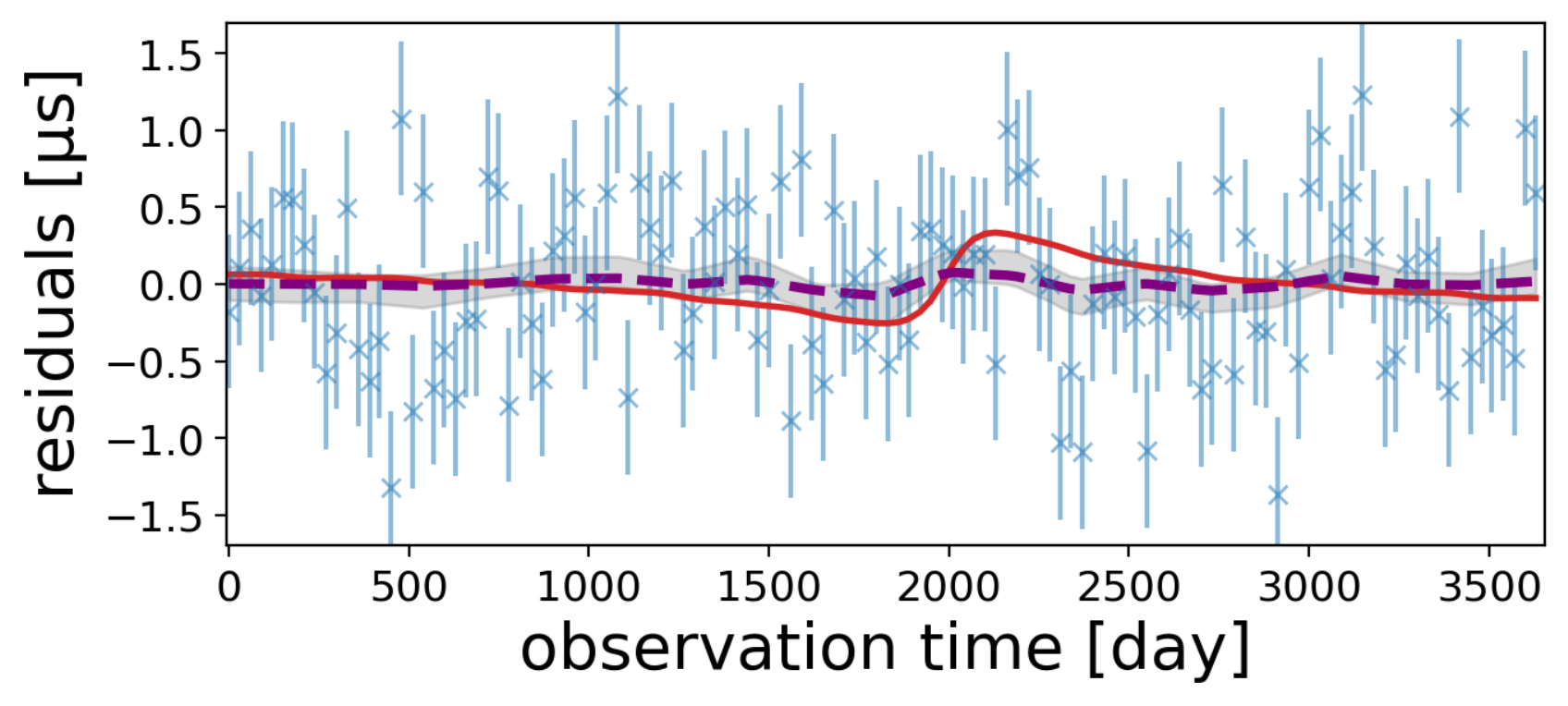}

\caption{\label{fig:PTA-1-1}Post-fit residuals (blue data points with error bars) and perturbations from the injected weak burst (solid red curves) for 4 of the 20 pulsars. {These are  the same pulsars as those in Fig. \ref{fig:PTA-1}. The difference in residuals for each pulsar comes from the distance of the encounter event.} {The reconstructed burst-induced residuals are shown as dashed, purple curves (with 90\% confidence intervals), where the sky location of the burst is taken as the  posterior medians for $\cos\theta$  and $\phi$.}}
\end{figure}

\subsection{Weak signal ($\text{SNR}\approx6.5$)}

\begin{figure}
\includegraphics[scale=0.5]{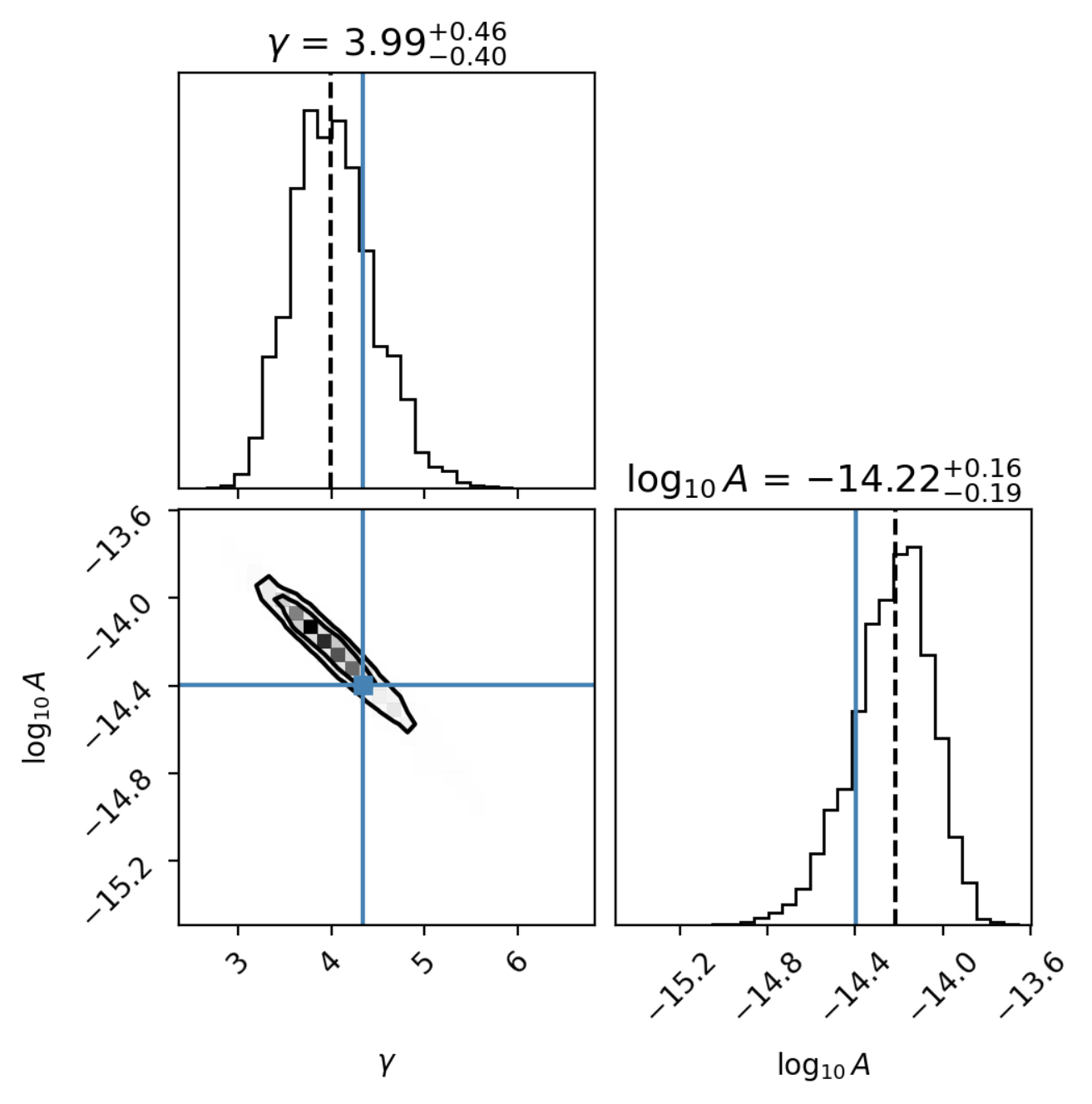}

\caption{\label{fig:null1-1}Posterior distributions of the SGWB parameters $A$ and $\gamma$ in the noise-only model when a weak burst signal is present.  The dashed lines represent the median values, and the blue solid lines denote the true values in the simulated data set. {For such a weak signal, the SGWB parameters are properly captured by our model.}}
\end{figure}

Now we turn to a weaker signal from the encounter event occurring at a distance of 45 Mpc. Fig. \ref{fig:PTA-1-1} shows the timing residuals and the contribution from the injected burst for 4 of the 20 pulsars (the same  as those in Fig. \ref{fig:PTA-1}). By analyzing the data set with the noise-only model, we obtain the posteriors shown in Fig. \ref{fig:null1-1}. Unlike the case in the previous subsection, the distributions of $\log_{10}A$ and $\gamma$ here are compatible with the true values. The burst signal is so weak that the noise-only model is capable of detecting the injected SGWB.

 \begin{figure*}
\includegraphics[scale=0.5]{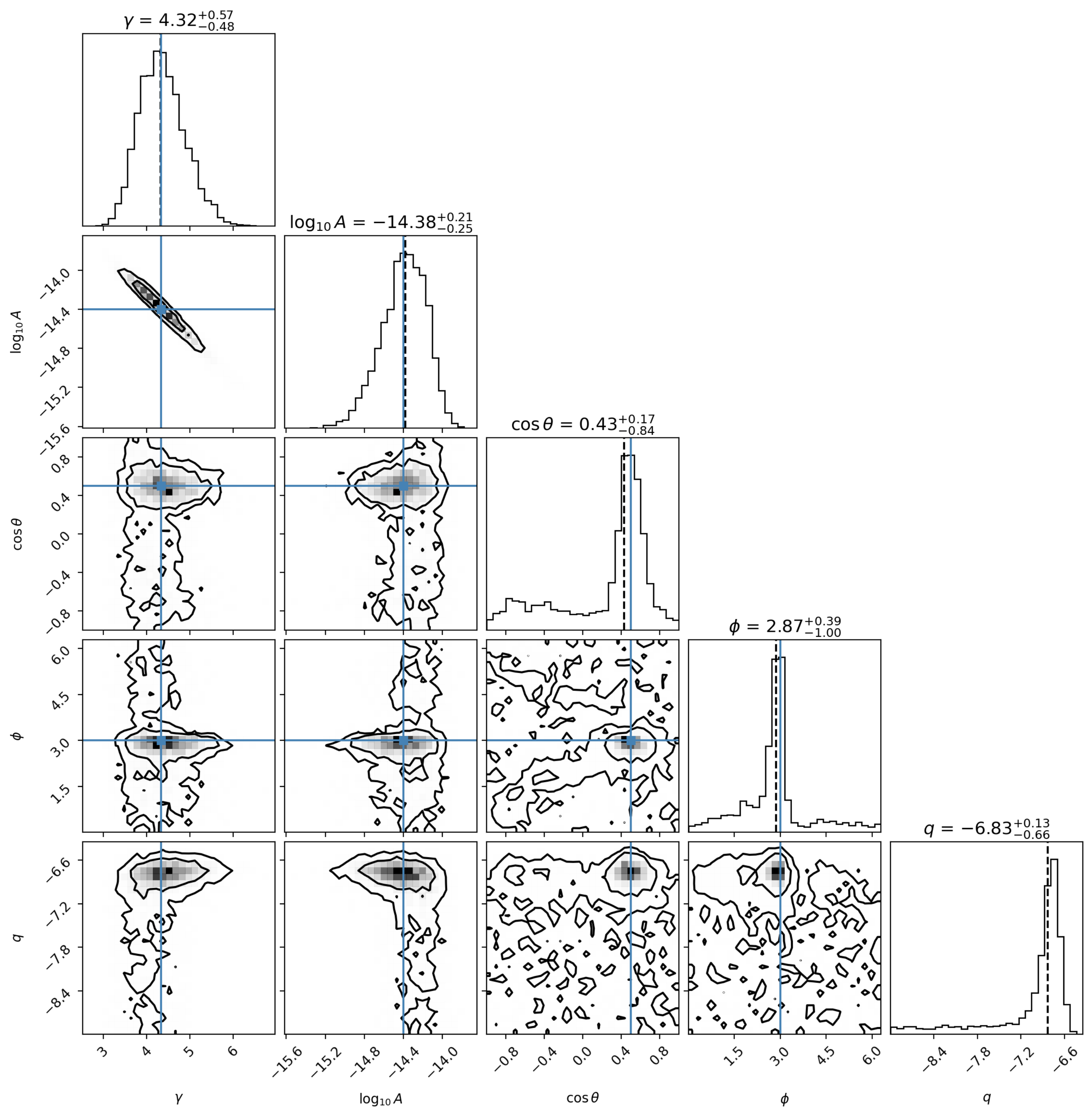}

\caption{\label{fig:model1-1}Posterior distributions of parameters $\log_{10}A,\gamma,\cos\theta,\phi$ and $q$ in our model when a weak signal is present. The dashed lines represent the median values, and the blue solid lines denote the true values in the simulated data set. {The red noise parameters are recovered and the location of the burst is detected. However, the evidence supporting the existence of the burst is not significant (see text).}}
\end{figure*}
\begin{figure}
\centering
\includegraphics[scale=0.45]{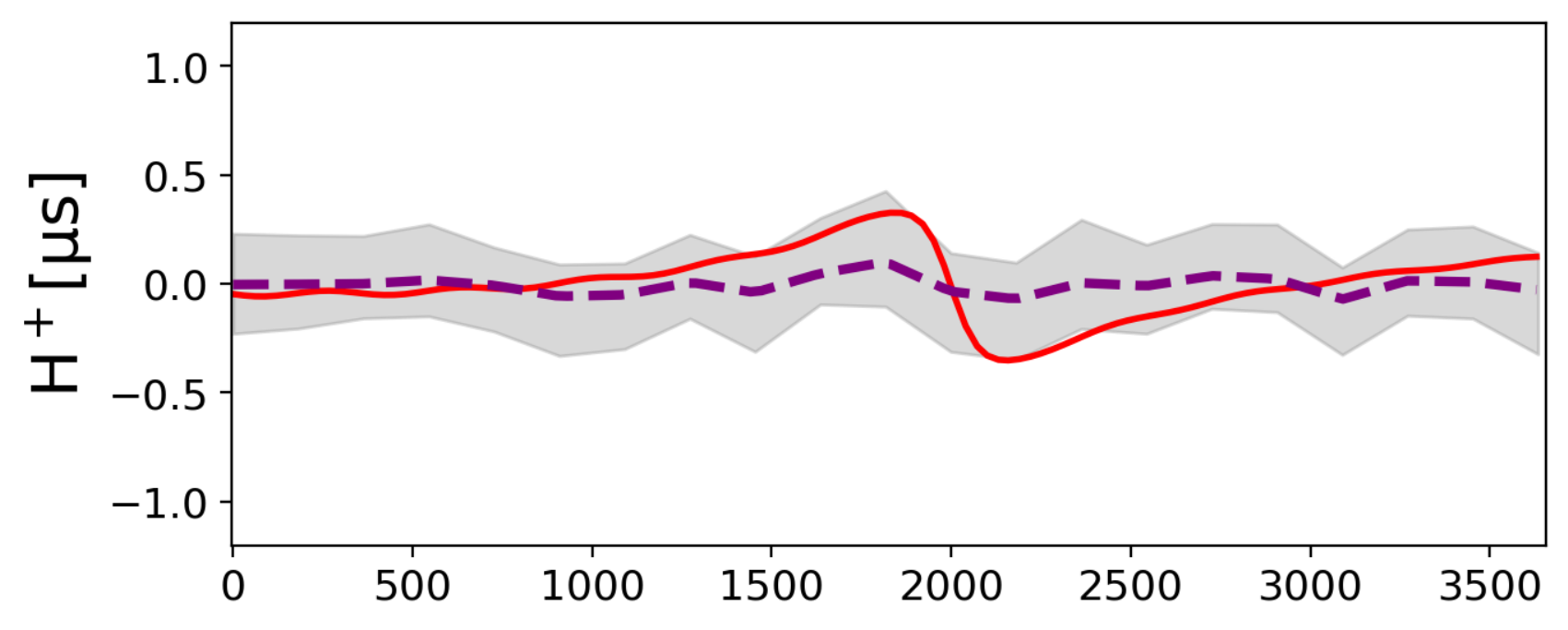}
\includegraphics[scale=0.45]{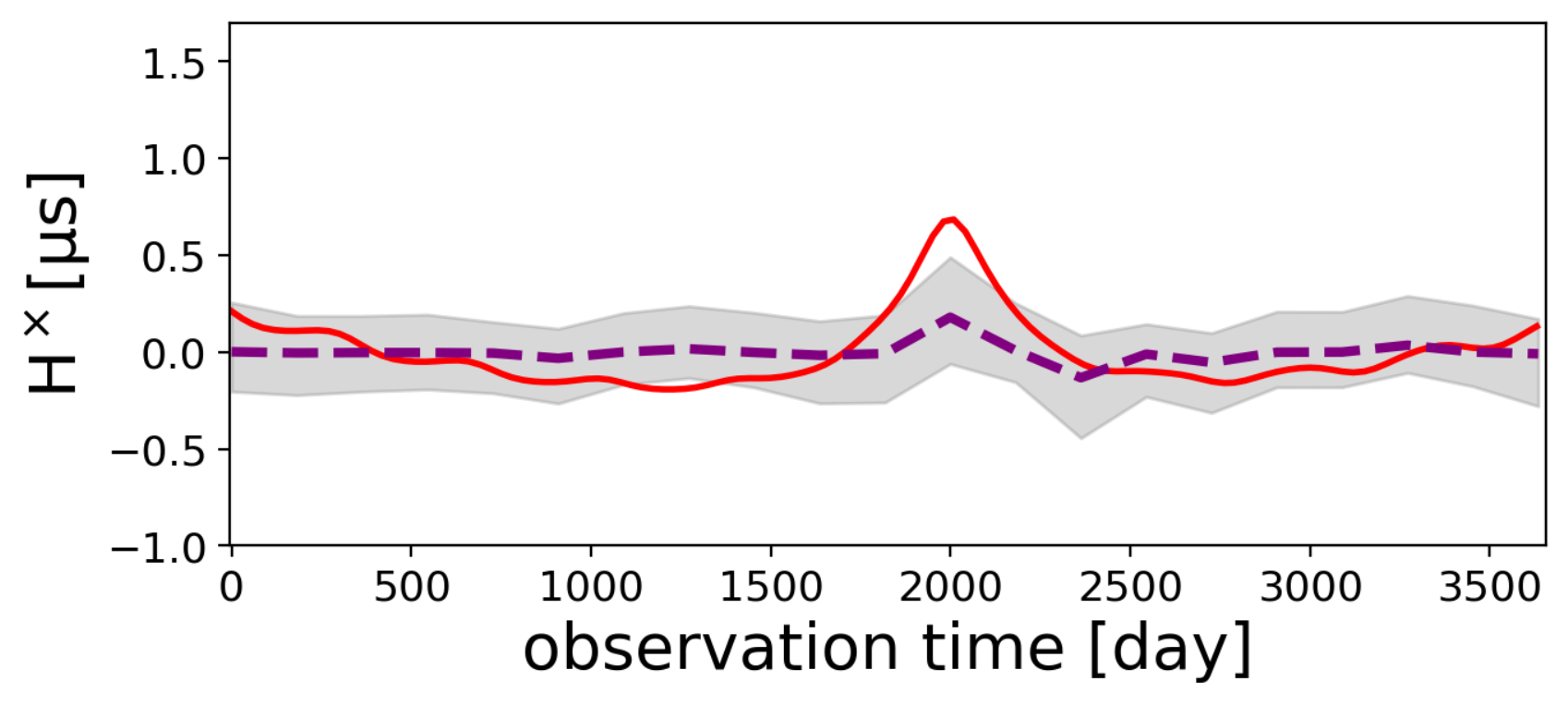}
\caption{\label{fig:waveform_reconstruction-1}Injected $H^{+}(t)$ and $H^{\times}(t)$ (post-fit) for a weak burst signal (red curves), the reconstructed piecewise linear functions $\mathsf{H}_{\mu}^{+}$ and $\mathsf{H}_{\mu}^{\times}$ (purple curves) and 90\% confidence intervals (shaded region), {where the sky location of the burst is taken as the posterior medians for $\cos\theta$  and $\phi$. For such a weak signal, our model is not able to characterize the burst waveform. }}
\end{figure}

We then search for the burst using our model with additional parameters $\theta,\phi$ and $q$. The posterior distributions are shown in Fig. \ref{fig:model1-1}. We can see that the SGWB spectrum is also well captured, with median values and true values almost overlapping. The distribution of $q$ has a tail extending to the lower bound of its prior. Both the Savage-Dickey ratio and the model evidences computed by Nestle give a Bayes factor of $\sim2\text{-}3$, corresponding to fairly weak evidence supporting our model. However, the sky location of the injected burst is captured by the two peaks in the posteriors of $\cos\theta$ and $\phi$. The reconstructed waveform is shown in Fig. \ref{fig:waveform_reconstruction-1}. {We also show in Fig. \ref{fig:PTA-1-1}  the reconstructed burst-induced residuals for 4 pulsars.} As expected, the injected signal is too weak to be fully characterized by our model. 

\subsection{No signal}

\begin{figure*}
\includegraphics[scale=0.5]{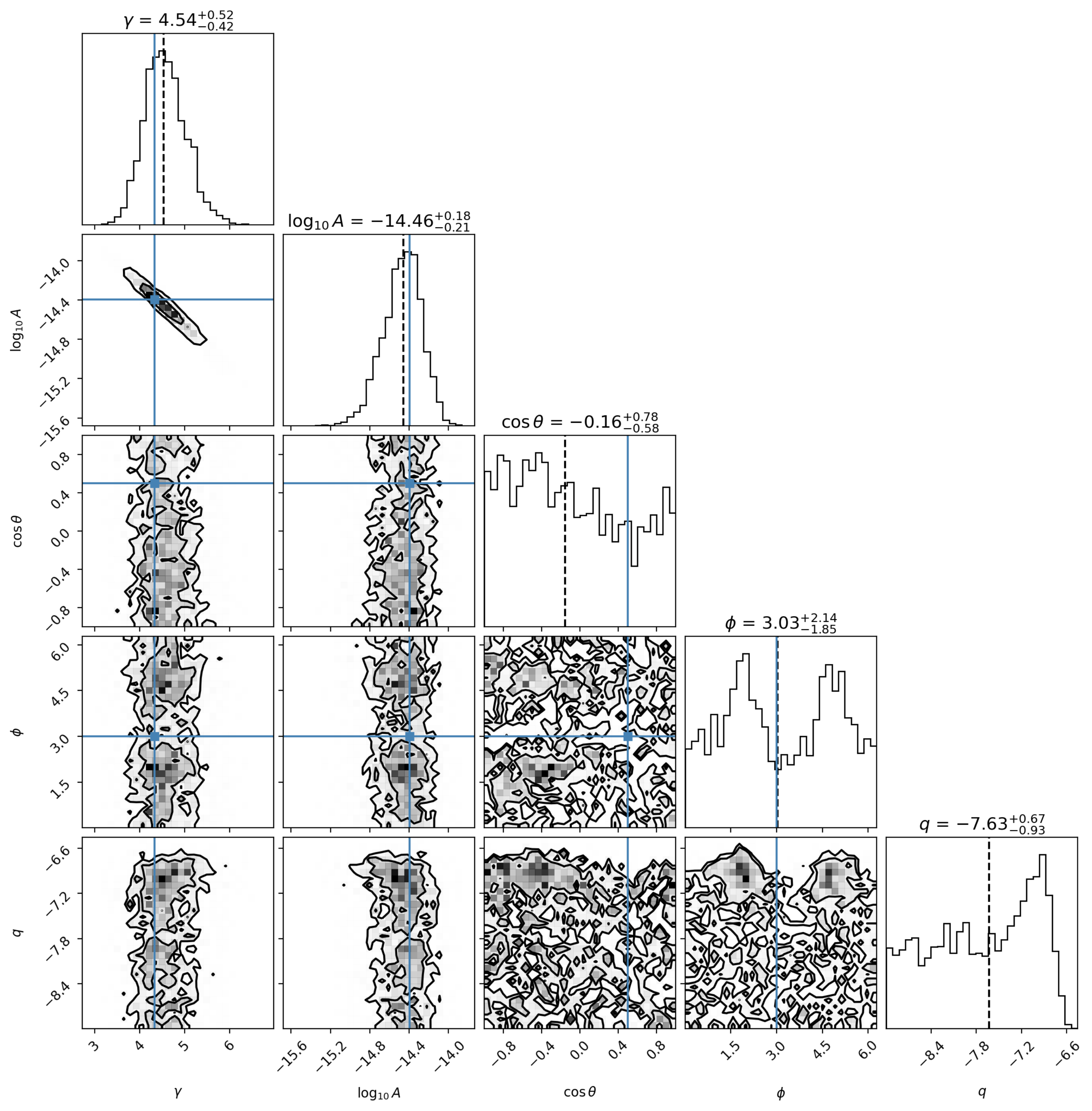}

\caption{\label{fig:model1-1-1}Posterior distributions of parameters $\log_{10}A,\gamma,\cos\theta,\phi$ and $q$ when no burst signal is present.  The dashed lines represent the median values, and the blue solid lines denote the true values in the simulated data set. {Our model does not detect any burst signal in this data set. The distributions of the SGWB parameters are compatible with the true values.}}
\end{figure*}

In the absence of deterministic signals, our model should be consistent with the noise-only model. Here we consider a data set without a burst.  The corner plots for our model are shown in Fig.  \ref{fig:model1-1-1}. The Bayes factor for our model vs. the noise-only model is  $\sim 1$, which means our model is capable of describing data sets that contain no evidence of a burst signal. 
 
\section{Conclusions and discussion \label{discussion}}
In this work, we have investigated a method for performing Bayesian analyses on PTA data sets to search for the strongest burst signals. The burst waveform is modeled by piecewise straight lines, which allows the likelihood to have a simple form. Although the number of waveform parameters $\boldsymbol{\mathsf{H}}$ could be large (of order $\mathcal{O}(10)$ depending on how the observation period is divided), these parameters can be analytically integrated out if their priors follow a Gaussian distribution. The resulting marginalized likelihood (Eq. (\ref{eq:version2})) has only three parameters in addition to the intrinsic and common noise parameters. Among the three parameters, $q$ controls the prior of the waveform parameters $\boldsymbol{\mathsf{H}}$; its marginal posterior can immediately tell us whether our model is favored over the noise-only model. The other two parameters, $\theta$ and $\phi$, denote the sky location of the burst. If a signal is present, one can efficiently retrieve the posterior of the waveform by analyzing the MCMC samples of $q, \theta$ and $\phi$ based on the marginalized likelihood.

We tested this model by analyzing three simulated PTA data sets, the first two containing a burst signal generated by the parabolic encounter of two SMBHs, and the third containing no burst signals. For the strong signal (with $\text{SNR}\approx 14.7$), our model is strongly favored compared to the noise-only model; not only can the burst's sky location be detected, its waveform can also be extracted to a reasonable accuracy. For a weak signal  (with $\text{SNR}\approx 6.5$), although the waveform cannot be distinguished from the background, the marginal posteriors of the sky location peak near the true values. When the signal is absent, the Bayes factor for our model to the noise-only model becomes $\sim 1$.

Our model could be improved in several ways. {As mentioned previously, in performing the piecewise linear fit, how the observation period is divided was predetermined. We fixed the grid point number $n_{\tiny H}$, with each segment containing a similar number of TOAs in each pulsar. In real data, however, TOAs are not evenly sampled in time, and may vary significantly from one pulsar to another. We expect that the optimal number of grid points and the optimal spacing will both depend on the signal. A higher SNR signal requires a larger $n_{\tiny H}$, and a signal highly concentrated in a particular time span requires more points in that region and fewer elsewhere. Noting that varying the grid spacing adaptively tends to significantly increase the computational cost, a straightforward extension of the current framework is to treat $n_{\tiny{H}}$ as a free parameter while ensuring we have a sufficient number of TOAs for each piece, or to test different $n_{\tiny{H}}$'s and then perform model selection on $n_{\tiny{H}}$.}

Finally, the burst may also be better characterized if the prior on the waveform parameters $\boldsymbol{\mathsf{H}}$ is not controled by a single hyperparameter. In the above analyses, we assume the prior $\pi(\boldsymbol{\mathsf{H}}|q)$ is a Gaussian distribution with covariance $\boldsymbol{Q}$ determined by $q$ only. {However, the amplitude of the waveform can differ between $H^{+}(t)$ and $H^{\times}(t)$, and} may vary significantly over the observation period. Roughly speaking, regions with a smaller amplitude require a smaller $q$. Hence a natural extension of the current framework is to set a number of hyperparameters (e.g., $q_1,q_2,...,q_n$, where $n\ll n_{\tiny H}$  in order not to lose the efficiency of the model) responsible {for different polarization components and} different time domains.

{We plan on improving the model accordingly and applying it to the search for a gravitational wave burst signal in real data sets in the near future.}

\section*{Acknowledgments}
This project receives support from National Science Foundation (NSF) Physics Frontiers Center award numbers 1430284 and 2020265,

\bibliography{PWLBurst}

\end{document}